\documentclass[final,5p,times,twocolumn]{elsarticle}

\usepackage{graphics}
\usepackage{stackengine}
\usepackage{multirow}
\usepackage[usenames]{color}
\usepackage[table]{xcolor}
\usepackage{makecell}
\usepackage[strings]{underscore}

\usepackage{amssymb}
\usepackage{amsmath}

\journal{Journal of Petroleum Science \& Engineering. Special Issue: Petroleum Data Science} 

\begin{document}

\begin{frontmatter}

\title{Data-driven model for hydraulic fracturing design optimization: focus on building digital database and production forecast}


\author[SKT]{A.D.~Morozov}
\author[SKT]{D.O.~Popkov}
\author[SKT]{V.M.~Duplyakov}
\author[SKT]{R.F.~Mutalova}
\author[SKT]{A.A.~Osiptsov}
\ead{a.osiptsov@skoltech.ru}
\author[SKT]{A.L.~Vainshtein}
\author[SKT]{E.V.~Burnaev}
\author[GPN]{E.V.~Shel}
\author[GPN]{G.V.~Paderin}

\address[SKT]{Skolkovo Institute of Science and Technology (Skoltech), 3 Nobel Street, 143026, Moscow, Russian Federation}
\address[GPN]{Gazpromneft Science \& Technology Center, 75-79 liter D Moika River emb., St Petersburg, 190000, Russian Federation}

\begin{abstract}
Growing amount of hydraulic fracturing (HF) jobs in the recent two decades resulted in a significant amount of measured data available for development of predictive models via machine learning (ML). In multistage fractured completions, post-fracturing production analysis (e.g., from production logging tools) reveals evidence that different stages produce very non-uniformly, and up to 30\% may not be producing at all due to a combination of geomechanics and fracturing design factors. Hence, there is a significant room for improvement of current design practices. We propose a data-driven model for fracturing design optimization, where the workflow is essentially split into two stages. As a result of the first stage, the present paper summarizes the efforts into the creation of a digital database of field data from several thousands of multistage HF jobs on vertical, inclined and near-horizontal wells from circa 20 different oilfields in Western Siberia, Russia. In terms of the number of points (fracturing jobs), the present database is a rare case of a representative dataset of about 5000 data points, compared to typical databases available in the literature, comprising tens or hundreds of points at best. Each point in the data base contains the vector of 92 input variables (the reservoir, well and the frac design parameters) and the vector of production data, which is characterized by 16 parameters, including the target, cumulative oil production. The focus is made on data gathering from various sources, data preprocessing and development of the architecture of the database as well as solving the production forecast problem via ML. Data preparation has been done using various ML techniques: the problem of missing values in the database is solved with collaborative filtering for data imputation; outliers are removed using visualisation of cluster data structure by t-SNE algorithm. The production forecast problem is solved via CatBoost algorithm. Prediction capability of the model is measured with the coefficient of determination ($R^2$) and reached 0.815. The inverse problem (selecting an optimum set of fracturing design parameters to maximize production) will be considered in the second part of the study to be published in another paper, along with a recommendation system for advising DESC and production stimulation engineers on an optimized fracturing design.
\end{abstract}

\begin{keyword}
bridging \sep fracture \sep particle transport \sep viscous flow \sep machine learning \sep predictive modelling \sep data collection \sep design optimization
\end{keyword}

\end{frontmatter}

\section{Introduction and problem formulation}


\subsection{Introductory remarks}
Hydraulic fracturing (in what follows referred to as HF for brevity) is one of the most widely-used techniques for stimulation of oil and gas production from wells drilled in the hydrocarbon-bearing formation~\cite{economides}. The technology is based on pumping at high pressures the fluid with proppant particles downhole through the tubing, which creates fractures in the reservoir formation. The fractures filled with granular material of closely packed proppant particles at higher-than-ambient permeability provide conductive channels for hydrocarbons from far-field reservoir to the well all the way to surface. The technology of HF began in 1947 as an experiment by Stanolind Oil with  gasoline as the carrier fluid and a sand from the Arkansas river at the Hugoton gas
field in Grant County, Kansas, US~\cite{HF-history}. In 1949, first commercial treatments were applied by Halliburton Oil Well cementing Company in Texas and Oklahoma, U.S. In the Soviet Union, the fracturing treatments were first performed in early 1950-s, on oil fields as well as for stimulation of coal bed methane production. Over the last two decades, the technical complexity of the stimulation treatment has made a significant step forward: wells are drilled directionally with a near-horizontal segment and multistage fractured completion. 

The global aim of this study is to structure and classify existing machine learning (ML) methods and to highlight the major trends for HF design optimization. Gradual development of fracturing technology is based on the advances in chemistry \& material science (fracturing fluids with programmed rheology, proppants, fibers, chemical diverters), mechanical engineering (ball-activated sliding sleeves for accurate stimulation of selected zones), and the success of fracturing stems from it being the most cost effective stimulation technique. At the same time, fracturing may be perceived as yet not fully optimized technology in terms of the ultimate production: up to 30\% of fractures in a multi-stage fractured completion are not producing~\cite{PLT1,PLT2}. For example, \cite{Miller2011} analyzed distributed production logs from various stages along the near-horizontal well and concluded that almost one third of all perforation clusters are not contributing to production. The reasons for non-uniform production from various perforation clusters along horizontal wells in a plug-and-perf completion are ranging from reservoir heterogeneity and geomechanics factors to fracturing design flaws. Thus, the pumping schedule has yet to be optimized, and it can be done either through continuum mechanics modeling (commercial fracturing simulators with optimization algorithms) or via data analytics techniques applied to a digital field database. We chose the latter route. To resolve this problem three initial classification categories are suggested: descriptive big data analytics should answer what happened during the job, predictive analytics should improve the design phase, and prescriptive analytics is to mitigate production loss from unsuccessful jobs. Here in Part I, we begin with the forward problem (predicting oil production rate from HF design and reservoir geology) in order to be able to solve the inverse problem in what follows in Part II. The question in phase II will be posed as: what is the optimum design of an HF job in a multistage fracturing completion to reach the highest possible ultimate cumulative production?

\subsection{Recent boom in shale fracturing} 

The boom in shale gas/shale oil fracturing owing to the simultaneous progress in directional drilling and multistage fracturing has resulted in extra supply in the world oil market, turning U.S. into one of the biggest suppliers. As a by-product of the shale gas technology revolution~\cite{king2010}, there is a large amount of high-quality digital field data generated by multistage fracturing operations in shale formations of the U.S. Land, that fuel the data science research into the HF design optimization~\cite{Shale-Analytics}. 

Modeling of shale reservoirs is a very comprehensive problem. The flow mechanism is not yet fully understood and properly simulated across the industry. The full scale simulation could be upgraded with ML-based pattern recognition technology where maps and history-matched production profile could enhance prediction quality for Bakken shale \cite{mohaghegh2011modeling}. Marcellus shale with similar approach is detailed in \cite{esmaili2016full}. Here the data driven analytics was used instead of classical hydrodynamic models.

Problem of activation of the natural fractures network by hydraulically-induced fractures is crucial for commercial production from this type of reservoirs. The mutual influence of natural and artificial fractures during the job has been studied by \cite{keshavarzi2013real}. The research has predicted the fracture behavior when it encounters a natural fracture with the help of Artificial Neural Network (ANN) \cite{Ensembles2013,HDA2013,ANNInit2016}. A similar approach is presented in \cite{guo2014new}. Three-level index system of reservoir properties evaluation is proposed to be a new method for gas well reservoir model control in fractured reservoir based on fuzzy logic theory and multilevel gray correlation.

In \cite{schuetter2015data} the authors developed a decision procedure to separate good wells from poor performers. For this purpose, the author investigated Wolfcamp well dataset. Analysis based on Decision Trees is applied to distinguish top quarter of wells from the bottom quarter. Most influential subset of parameters, characterizing a well, is also selected. 

\subsection{Prior art in frac design and its optimization}
\label{sec13}
Typically the oilfield services industry is using numerical simulators based on the coupled solid-fluid mechanics models for evaluation and parametric analysis of the HF job~\cite{detournay2016mechanics, osiptsov2017review, economides2002unified}. 
There is a variety of HF simulators based on KGD, PKN, P3D, or Planar3D models of the hydraulic fracture propagation process. Shale fracturing application called for more sophisticated approaches to modeling of the fracture network propagation. A good overview of the underlying models can be found in~\cite{detournay2016mechanics, osiptsov2017review}. Once there is a robust forward model of the process, an optimization problem can be posed with a prescribed objective function~\cite{queipo2002GlobOptHF}. Particular case of stimulation in carbonate reservoirs is acid frac. Iranian field with 20 fractured well has been studied by \cite{zoveidavianpoor2012development} in order to test candidate selection procedure.

A typical approach to the optimization problem includes the construction of a surrogate (see \cite{GTApprox2016}) of an objective function, whose evaluation involves the execution of a HF simulator. The computational
model integrates a hydraulic fracture simulator to predict propped fracture geometry and a production model to estimate the production flow rate. Then, an objective function is calculated, which can be any choice from papers listed in Section \ref{sec13} above. An example of the realization of such optimization strategy is presented in detail in~\cite{queipo2002GlobOptHF}. Another example of an integrated multiobjective optimization workflow is given in~\cite{rahman2001Multi-Objective-Opt}, which involves a coupling of the fracture geometry module, a hydrocarbon production module and an investment-return cash flow module.

\subsection{ML for frac design optimization}

In North America, thanks to the great attention to multistage fracturing in shales there is an increasing amount of research papers studying the application of big data analytics to the problem of HF optimization. 

A general workflow of the data science approach to HF for horizontal wells implicate techniques that cluster similar critical time-series into Frac-Classes of frac data (surface treatment pressure, slurry pumping rates, proppant loading, volume of proppant pumped). Correlation of the average Frac-Classes with 30-day peak production is used on the second step to distinguish between geographically distinct areas~\cite{anderson2016using}.

Statistically representative synthetic data set is used occasionally to build data-driven fracture models. The performance of the data-driven models is validated by comparing the results to a numerical model. Parameters include the size, number, location, and phasing angle of perforations, fluid and proppant type, rock strength, porosity, and permeability. Data-driven predictive models (surrogate models, see \cite{GTApprox2016,surrmod}) are generated by using ANN and Support Vector Machine (SVM) algorithms~\cite{temizel2015efficient}. Another approach to constructing metamodels on transient data (time series) is Dynamic Mode Decomposition (DMD), which is being explored, e.g., in~\cite{meta2020}.

Important geomechanics parameters are Young's modulus and Poisson's ratio obtained from geomechanics lab tests on core samples, that is far away from covering full log heterogeneity with missing values, hence the authors used Fuzzy Logic, Functional Networks and ANNs \cite{abdulraheem2009prediction}.

A detailed literature review on the subject of frac design optimization is provided by \cite{gao2017design}, where the authors emphasized the necessity of bringing a common integrating approach into the full scale on shale gas systems. The data-driven analytics was proposed as a trend in the HF design optimization. Authors induced game-theoretic modeling and optimization methodologies to address multiple issues. The impact of proppant pumping schedule during the job has been investigated in \cite{poulsen1986procedure} by coupling fractured well simulator results and economical evaluations.  

There are several approaches with different target criteria for optimization. For a wide variety of reasons, the proppant fraction is quite an important parameter to evaluate. In \cite{saldungaray2012hydraulic}, the authors reviewed four major case studies based on shale reservoirs across the U.S. and suggesting strategy to evaluate the realistic conductivity and impact on stimulation economics of proppant selection. 

Field data, largely accumulated over the past decades, are being digitized and structured within oil companies. The market landscape in the era of declining oil prices after 2014 has stimulated shale operators to look closer at the capabilities of data science to optimize the fracturing technology \cite{betz2015low}. The issue of working with short-term data and the need to find a way to turn that into long-term well performance was emphasized. Proppant loading was shown to be one of the most important variables for productivity. Increasing industry interest to artificial intelligence and to application of ML algorithms is justified by the combination of several factors: processing power growth and amount of data available for analysis. Thousands of completions are digitized~(e.g., see~\cite{awoleke2011analysis}), giving the grounds for the use of a wide range of big data analytics methods. One of the most recent studies~\cite{wang2019insights} investigated the relationships between the stimulation parameters and first-year oil production for a database of horizontal multistage fractured wells drilled in unconventional Montney formation in Canada. Four commonly used supervised learning approaches including Random Forest (RF), AdaBoost, SVM, and ANN \cite{Hastie2010} were evaluated to demonstrate that the RF performs the best in
terms of prediction accuracy.  

The state of affairs is a bit different in other parts of the world, where, though the wells are massively fractured, the data is not readily available and is not of that high quality as in the North America Land, which poses a known problem of ``small data'' analysis, where neural networks do not work, and different approaches are called for. 

In Russia, there are a few attempts of using ML algorithms to process data of HF, e.g., the paper~\cite{alimkhanov2014application} presents the results of developing a database of 300 wells, where fracturing was performed. Operational parameters of the treatments were not taken into account in this paper. Classification models were developed to distinguish between efficient/inefficient treatments. Job success criteria were suggested in order to evaluate the impact of geological parameters on the efficiency via classification. Regression models were proposed for predicting post-frac flow rate and water cut. A portfolio of standard algorithms was used such as decision tree, random forest, boosting, ANNs, linear regression and SVM. Limitations of linear regression model applied for water cut prediction were discussed. Recent study~\cite{makhotin2019gradient} used gradient boosting to solve the regression problem for predicting the production rate after the simulation treatment on a data set of 270 wells. Mathematical model was formulated in detail, though data sources and the details of data gathering and preprocessing were not discussed.

\subsection{Problem Statement}
To summarize the introductory remarks presented above, HF technology is a complex process, which involves accurate planning and design using multi-discipline mathematical models based on coupled solid~\cite{detournay2016mechanics} and fluid~\cite{osiptsov2017review} mechanics. At the same time, the comparison of flow rate prediction from reservoir simulators using fracture geometry predicted by HF simulators vs. real field data suggests there is still significant uncertainty in the models. The two step model of fracturing and production is being extended to include the transient flowback into the integrated optimization workflow~\cite{osiptsov2020flowback}, but in the present study we focus on HF design only, leaving flowback optimization based on data analysis for a separate study.  


In contrast to the traditional methodology of making the design of fracturing technology based on parametric studies with an HF simulator, we propose to investigate the problem of design optimization using ML algorithms on field data from HF jobs, including reservoir, well, frac design, and production data. As a training field database, we will consider the real field data collected on fracturing jobs in Western Siberia, Russia. 

The entire database from real fracturing jobs can be conventionally split into the input data and the output data. The input data, in turn, consists of the parameters of the reservoir and the well (permeability, porosity, hydrocarbon properties, etc.) and the frac job design parameters (pumping schedule). The output is a vector of parameters characterising production. 

The usefulness of hybrid modeling is widely reported in the literature \cite{helmy2010hybrid}. Numerous efforts have been made by researches to implement data science to lab cost reduction issues. PVT correlations correction for crude oil systems were comparatively studied between ANN and SVM algorithms~\cite{el2007support}.  

Finally, the problem at hand is formulated as follows: one may suppose that a typical hydraulically-fractured well does not reach its full potential, because the fracturing design is not optimum. Hence, a scientific question can be posed within the big data analysis discipline: what is the optimum set of fracturing design parameters, which for a given set of the reservoir characterization-well parameters yield an optimum post-fracturing production (e.g., cumulative production over a given period, say 3 months)? It is proposed to develop a ML algorithm, which would allow one to determine the optimum set of HF design parameters based on the analysis of the reservoir-well-flow rate data.

Out of this study we expect also to be able to make recommendations on
\begin{itemize}
    \item[---] oil production forecast based on the well and the reservoir layer data;
    \item[---] the optimum frac design;
    \item[---] data acquisition systems, which are required to improve the quality of data analytics methods.
\end{itemize}

In the course of the study we will focus on checking the following hypotheses and research questions: 
\begin{enumerate}
\item Is there a systematic problem with HF design?
\item What is the objective function for optimization of HF design? What are various metrics of success?
\item How do we validate the input database?
\item What database is full (sufficient)? (Optimum ratio of number of data points vs. number of features for the database?)
\item What can be learned from field data to construct a predictive model and to optimize the HF design?
\item Is there a reliable ML-based methodology for finding the optimum set of parameters to design a successful HF job?
\end{enumerate}

At the first stage of the entire workflow, we are aimed at collecting a self-consistent digital data base of several thousand data points (each containing infromation about the reservoir, well and frac design parameters) and solving the production forecast problem with ML methods. At the second stage, we will consider the inverse optimization problem.

\subsection{Metrics of success for a fracturing job}
The ultimate optimization of a stimulation treatment is only possible if the outcome is measured. Below we summarize various approaches to quantify the success of an HF job: 
\begin{itemize}

\item Cumulative oil production of 6 and 18 months is used by \cite{wang2016comprehensive} as a target parameter, and is predicted by a model with 18 input parameters, characterizing Bakken formation in North America.   

\item Predictive models for the 12 months cumulative oil production  are built by \cite{schuetter2018data} using multiple input parameters characterizing well location, architecture, and completions.

\item Feed-forward neural network was used by \cite{awoleke2011analysis} to predict average water production for wells drilled in Denton and Parker Counties, Texas, of the Barnett shale based on average monthly production. The mean value was evaluated using the cumulative gas produced normalized by the production time.

\item In \cite{balen1988applications}, a procedure was presented to optimize the fracture treatment parameters such as fracture length, volume of proppant and fluids, pump rates, etc. Cost sensitivity study upon well and fracture parameters vs NPV as a maximization criteria is used. Longer fractures does not necessarily increase NPV, a maximum discounted well revenue is observed by \cite{hareland1993hydraulic}.

\item Statistically representative set of synthetic data served as an input for ML algorithm in~\cite{temizel2015efficient}. The study analyzed the impact of each input parameter to the simulation results like cumulative gas production for contingent resources like shale gas simulation model. 

\item  $\Delta Q = (Q_2-Q_1)$ was an uplift metric to seek the re-fracture candidate for 50 wells oilfield dataset using ANN to predict after the job oil production rate $Q_2$ based on $Q_1$ oil production rate before the job \cite{yanfang2014refracture}.

\item  $Q$ pikes approach is presented by implementing B1, B2 and B3  statistical moving average for one, three and twelve-month best production results consequently in \cite{pankaj2018application}. The simulation is done over 2000 dimension dataset to reap the benefit from proxy modeling treatment.

\item Net present value is one of the metrics used to evaluate the success of a HF job~\cite{economides1988NPV}. Economical bias for HF is detailed by \cite{balen1988applications}. The proposed sequential approach of integrating upstream uncertainties to NPV creates an important tool in the identification of the crucial parameters affecting a particular job.

\end{itemize}

In Table \ref{my-label}, we compose a list of the main metrics for evaluation of HF job efficiency.

\begin{table}[]
\begin{tabular}{|l|l|}
\hline
\multicolumn{1}{|c|}{Metrics} & \multicolumn{1}{c|}{Source} \\ \hline
\begin{tabular}[c]{@{}l@{}}Cumulative oil production 6/18 month \\ just after the job\end{tabular}              & {\cite{wang2016comprehensive}}                    \\ \hline
12 months cumulative oil production                                                                    & {\cite{schuetter2018data}}                     \\ \hline
Average monthly oil production after the job                                                                    & {\cite{awoleke2011analysis}}                     \\ \hline
NPV                                                                                                         & {\cite{balen1988applications}}                     \\ \hline
Comparison to modelling                                                                                     & {\cite{temizel2015efficient}}                     \\ \hline
Delta of averaged Q oil                                                                                     & {\cite{yanfang2014refracture}}                     \\ \hline
\begin{tabular}[c]{@{}l@{}}Pikes in liquid production for 1, 3 \\ and 12 months\end{tabular}                & {\cite{pankaj2018application}}                     \\ \hline
\begin{tabular}[c]{@{}l@{}}Break even point (job cost equal to \\ total revenue after the job)\end{tabular} & {\cite{alimkhanov2014application}}                     \\ \hline
\end{tabular}
\caption{Success metrics of HF job}
\label{my-label}
\end{table}

\section{Overview of ML methods used for HF optimization}

ML is a broad subfield of artificial intelligence aimed to enable machines to extract patterns from data based on mathematical statistics, numerical methods, optimization, probability theory, discrete analysis, geometry, etc. ML tasks are the following: classification, regression, dimensionality reduction, clustering, ranking and others. Also, ML is subdivided into supervised/unsupervised and reinforcement learning.

Supervised ML problem can be formulated as constructing a target function $\hat{f}: X \rightarrow Y$ approximating $f$ given a learning sample $S_{m}=\{(x_{m},y_{m})\}$, where $x_m \in X$, $y_m \in Y$ with $y_i=f(x_i)$.

To avoid overfitting (discussed in the next section), it is also very important to select ML model properly. This choice largely depends on the size, quality and nature of the data, but often without a real experiment it is very difficult to answer which of the algorithms will be really effective.

The lack of data becomes one of the most common problems when dealing with field data. Some ML models can manage it (decision trees), while others are very sensitive to sparse data (ANNs). A number of the most popular algorithms such as linear models or ANNs do not cope with the lack of data; SVMs have a large list of parameters that need to be set, and the trees are prone to overfitting.

In our work, we want to show how strongly the choice of the model and the choice of the initial sample can affect the final results and the correct interpretation.

Actually, there are articles with results on application of ML to HF data that describe models with high predictive accuracy. However, the authors use small samples with rather homogeneous data and complex models prone to overfitting. We claim that more investigations are needed, evaluating prediction accuracy and stability separately for different fields and types of wells.

\subsection{Overfitting}
\label{overfitting}
Nowadays there exists an increasing number of papers about application of ML in HF data processing. However, many studies may rise questions on the validity of results in light of potential overfitting due to small data involved.

\begin{figure}[h!]
\includegraphics[width=8cm]{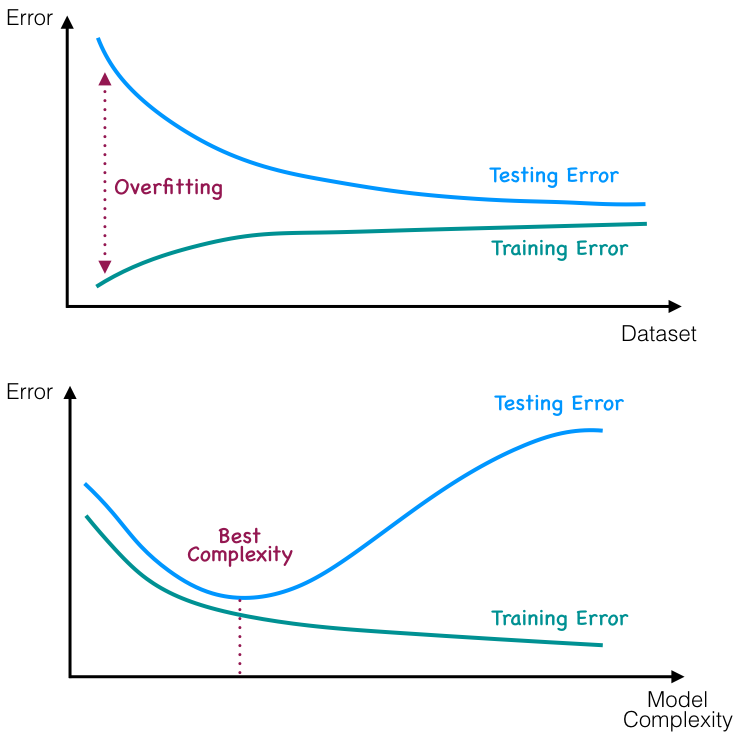}
\caption{Overfitting}
\label{Fig1}
\end{figure}

Overfitting is a negative phenomenon that occurs when the learning algorithm generates a model that provides predictions mimicking a training dataset too accurately, but have very inaccurate predictions on the test data \cite{Hastie2010}. In other words, overfitting is the use of models or procedures that violate the so-called Occam Razor \cite{hawkins2004problem}: the models include more terms and variables than necessary, or use more complex approaches than required. Figure \ref{Fig1} shows how the pattern of learning on test and training datasets changes dramatically, if overfitting takes place.

There are several reasons for this phenomenon \cite{hawkins2004problem, baumes2006support}:
\begin{itemize}
    \item Traditional overfitting: training a complex model on a small amount of data without validation. This is a fairly common problem, especially for industries that not always have access to big datasets, such as medicine, due to the difficulties with data collection. 
    \item Parameter tweak overfitting: use a learning algorithm with many parameters. Choose the parameters based on the test set performance.
    \item Bad statistics: misuse statistics to overstate confidence. Often some  known-false assumptions about some system are made and then excessive confidence of results is derived, e.g., we use Gaussian assumption when estimating confidence.
    \item Incomplete prediction: use an incorrectly chosen target variable or its incorrect representation, e.g. there is a data leak and inputs already contain target variable.
    \item Human-loop overfitting: a human is still a part of the learning process, he/she selects hyperparameters, creates a database from measurements, so we should take into account overfitting by the entire human/computer interaction.
    \item Dataset selection: purposeful use of data that is well described by the models built. Or use an irrelevant database to represent something completely new. 
    \item Overfitting by review: if data can be collected from various sources, one may select only the single source due to economy of resources for data collection, as well as due to computational capabilities. Thus, we consciously choose only one point of view.
\end{itemize}

For example, in the article \cite{alimkhanov2014application} only 289 wells, each described by 178 features, were considered for the analysis. This number of points is too small compared to the number of input features, so a sufficiently complex predictive model simply ``remembers'' the entire dataset, but it is unlikely that the model is robust enough and can provide reliable predictions. This is also confirmed by a very large scatter of results: the coefficient of determination varies from 0.2 to 0.6.

In this context you can find many articles, which used small data, of the order of 100 data points (150 wells were considered in \cite{mohaghegh2002identification}, 135 wells in \cite{esmaili2016full}, etc.). In addition, each of the mentioned studies uses a very limited choice of input features, which exclude some important parameters of HF operation. For example, the work~\cite{lolon2016evaluating} uses the following parameters to predict the quality of the HF performed: stage spacing, cemented, number of stages, average proppant pumped, mass of liquid pumped, maximum treatment rate, water cut, gross thickness, oil gravity, Lower Bakken Shale TOC, Upper Bakken Shale TOC, total vertical depth. This set of parameters does not take into account many nuances, such as the geomechanical parameters of the formation or the completion parameters of the well. Quite good results were shown in \cite{schuetter2015data}; various models were considered, but it was noted that out of 476 wells, only 171 have records have no NaN values.

In addition to the problems described above, overfitting may be caused by using too complex models: in many articles they use one of the most popular ML methods, the artificial neural network (ANN). However, it is known that a neural network is a highly non-linear model that very poorly copes with the lack of data and is extremely prone to overfitting. Lack of data is a fairly frequent case when it comes to real field data, which makes the use of ANNs unreliable.

There are examples of using the SVM algorithm \cite{xiaofeng2016post}. The main disadvantage of SVM is that it has several key hyperparameters that need to be set correctly to achieve the best classification results for each given problem. 
The same hyperparameters can be ideal for one task and not fit at all for another. Therefore, when working with SVM a lot of experiments should be made, and the calculation takes a fairly large amount of time. Moreover, a human-loop overfitting can occur. The above algorithms work very poorly with missing values. 

In conclusion, to reduce overfitting and to construct a robust predictive model, the necessary condition is to develop a big and reliable training dataset that contains all required input features.

\subsection{Dimensionality reduction}
When a dataset has a large number of features (large dimension), it can lead to a large computation time and to difficulties in finding a good solution due to excessive noise in data. In addition, for larger feature dimension we need more examples in the data set to construct a reliable and accurate predictive model. In addition, a large dimension greatly increases the likelihood that two input points are too far away, which, like in case of outliers, leads to overfitting. Therefore, in order to decrease the input dimension and at the same time to keep the completeness of information with decreasing dimension, we can use special dimension reduction and manifold learning methods, see \cite{DRreg2,DRreg}. Lastly, dimensionality reduction helps visualizing multidimensional data. In our work, we will use the T-distributed Stochastic Neighbor Embedding (t-SNE) algorithm \cite{tsne} for visualization after dimensionality reduction and missing values imputation. 

\subsection{Clustering}
\label{clustering}

Clustering methods \cite{Hastie2010} are used to identify groups of similar objects in multivariate datasets. In other words, our task is to select groups of objects as close as possible to each other, which will form our clusters by virtue of the similarity hypothesis. The clustering belongs to the class of unsupervised learning tasks and can be used to find structures in data. Since our database includes 23 different oilfields, horizontal and vertical wells, as well as different types of fracture design, it would be naive to assume that data is homogeneous and  can be described by a single predictive model.

Thus, by dividing dataset in clusters we can obtain more homogeneous subsamples, so that ML algorithms can easily construct more accurate models on subsamples \cite{Grihon2013}. In addition, clustering can be used for detecting outliers \cite{outlier2,outlier} in a multidimensional space. We utilise this for further analysis. In our case, we used t-SNE to visualize a low-dimensional structure of the data set to extract clusters and identify outlying measurements.

\subsection{Regression}
\label{regression}
After selecting a specific sample of data, it is necessary to solve the regression problem, i.e., to restore a continuous target value $y$ from the original input vector of features $x$ \cite{surrogate,GTApprox2016}. The dependence of the mean value $\mu = f(x)$ of $y$ on $x$ is called the regression of $y$ on $x$.

In open literature, some authors considered different approaches how to define a target variable. In particular, cumulative production for 3, 6 and 12 months was taken as a target. However, we noted a strong correlation between values of cumulative production for 3, 6 and 12 months. Thus, as a target variable we consider values of cumulative production for 3 months because the production over a longer period of time is not always known and 3 months period is necessary and sufficient.

Once the regression model is built, we assess its accuracy on a separate test sample. As a prediction accuracy measure, we use the coefficient of determination. The coefficient of determination ($R^2$ --- $R$-squared) is the fraction of the variance of the dependent variable explained by the model in question.

\subsection{Ensemble of models}
The ensemble of models \cite{Hastie2010,Ensembles2013} uses several algorithms in order to obtain better prediction efficiency than could be obtained from each trained model individually.

Ensembles are very prone to overfitting due to their high flexibility, but in practice, some assembly techniques, such as bagging, tend to reduce overfitting. The ensemble method is a more powerful tool compared to stand-alone forecasting models, since it minimizes the influence of randomness, averaging the errors of each basic model and reduces the variance.

\subsection{Feature importance analysis}
The use of tree-based models makes it easy to identify features that are of zero importance, because they are not used when calculating prediction. Thus, it is possible to gradually discard unnecessary features, until the calculation time and the quality of the prediction becomes acceptable, while the database does not lose its information content too much.

There is the Boruta method \cite{JSSv036i11} which is a test of the built-in solutions for finding important parameters. The essence of the algorithm is that features are deleted that have a Z-measure less than the maximum Z-measure among the added features at each iteration. Also, the Sobol method \cite{sobol} is widely used for feature importance analysis. The method is based on the representation of the function of many parameters as the sum of functions of a smaller number of variables with special properties.

In addition, testing and verifying feature importance may be done with the one-variable-at-a-time (OVAT) method \cite{ovat}. It is a method of creating experiments involving testing of parameters one at a time instead of multiple factors simultaneously. It is primarily used when data is noisy and it is not obvious which features affect the target.

\subsection{Hyperparameter search}
Hyperparameter optimization is the problem of choosing a set of optimal hyperparameters for a learning algorithm. Whether the algorithm is suitable for the data directly depends on hyperparameters, which directly influence overfitting or underfitting. Each model requires different assumptions, weights or training speeds for different types of data under the conditions of a given loss function.

The most common method for optimizing hyperparameters is a grid search, which simply does a full search on a manually specified subset of the hyperparameter space of the training algorithm. Before using the grid search, a random search can be used to estimate the boundaries of a region, where parameters are selected. Moreover, according to the Vapnik-Chervonenkis theory, the more complex a model is, the worse its generalizing ability. Therefore, it is very important to select the model complexity conforming to the given data set, otherwise prediction will be unreliable. To check the generalization ability, we can use a cross-validation procedure.

\subsection{Uncertainty Quantification}
Uncertainty comes from errors of the ML algorithm and from noise in the data set. Hence, predicting an output only is not sufficient to be certain with results. Therefore we should also quantify  uncertainty of the prediction. This can be done by using prediction intervals providing probabilistic upper and lower bounds on an estimate of the output variable. 


The prediction interval depends on some combination of the estimated variance of the model and the variance of the output variable caused by noise. 
The variance of the model is due variance of model parameters estimates, resulted from noise in the original data set. By building confidence intervals for the parameters and propagating them through the model we can estimate the variance of the model. In practice, to build prediction interval for a general nonlinear model we can use the bootstrap resampling method, although it is rather computationally demanding \cite{Hastie2010}.

Let us note that the difference between prediction and confidence intervals: the former quantifies the uncertainty on a single observation, estimated from the population, and the latter quantifies the uncertainty on an estimated population variable, such as a mean or a standard deviation. Let us note that it is important to quantify uncertainty on ML model performance, which we can do by estimating the corresponding confidence intervals.


Besides prediction or confidence intervals, another important type of uncertainty quantification is related to forward uncertainty propagation when we estimate how the variability of input parameters affects the output variance of the model. This helps to select the most important input features \cite{sobol,ADoESobol2017}.





\begin{figure*}[h]
\centering
\includegraphics[width=15cm]{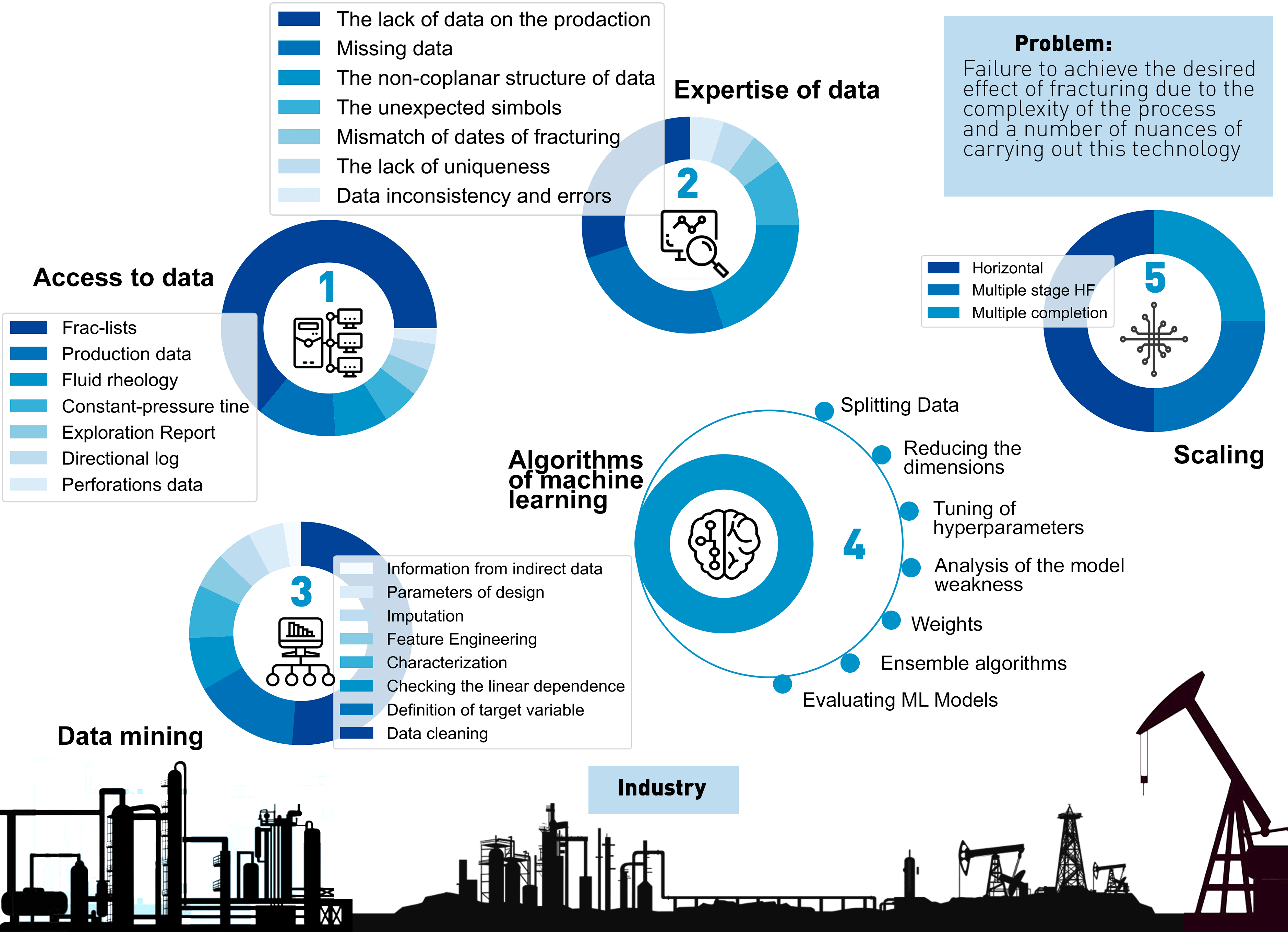}
\caption{General workflow}
\label{Fig1a}
\end{figure*}

\section{Field database: structure, sources, pre-processing, statistical properties, data mining}

Following the report by McKinsey\&Company from 2016 the majority of companies get real profit from annually collected data and analytics~\cite{mckinsey}. However, the main problem companies usually face while getting profit from data lies inside the organizational part of the work.

Most of the researches skip the phase of data mining, considering the ready-made dataset as a starting point for ML. Nevertheless, we can get misleading conclusions from false ML predictions due to learning on the low-quality dataset. As follows from results of \cite{gaurav2017horizontal} the most important thing when doing the ML study is not only a representative sample of the wells, but also a complete set of parameters that can fully describe the fracture process with the required accuracy.

As can be seen from Section \ref{overfitting}, where we describe various types of overfitting, the key issue is related to poor quality of the training dataset. In addition, if in case of a non-representative training dataset we use a subsample to train the model, corresponding results will be very unstable and will hide the actual state of affairs.

It is known that data pre-processing actually takes up to 3/4 of the entire time in every data-based project \cite{cleandata}. Having a good, high-quality and validated database is the key to obtain the interpretable solution using ML. The database must include all the parameters that are important from the point of view of the physics of the process, be accurate in its representation and be verified by subject domain experts in order to avoid the influence of errors in database maintenance.

Unfortunately, in field conditions each block of information about different stages of the HF is recorded in a separate database. As a result, there is no integrated database containing information about sufficient number of wells that would include all factors for decision making. So, we should first develop a correct procedure for data preprocessing in order to make a given data set more useful, work -- more efficient, results -- more reliable.

In the following subsections, we describe in detail the steps of forming the database, prior to applying ML algorithms. The entire workflow of the study with indication of different phases of development is shown in Figure \ref{Fig1a}.

\subsection{Collecting the database}

\begin{figure}[h!]
\includegraphics[width=8cm]{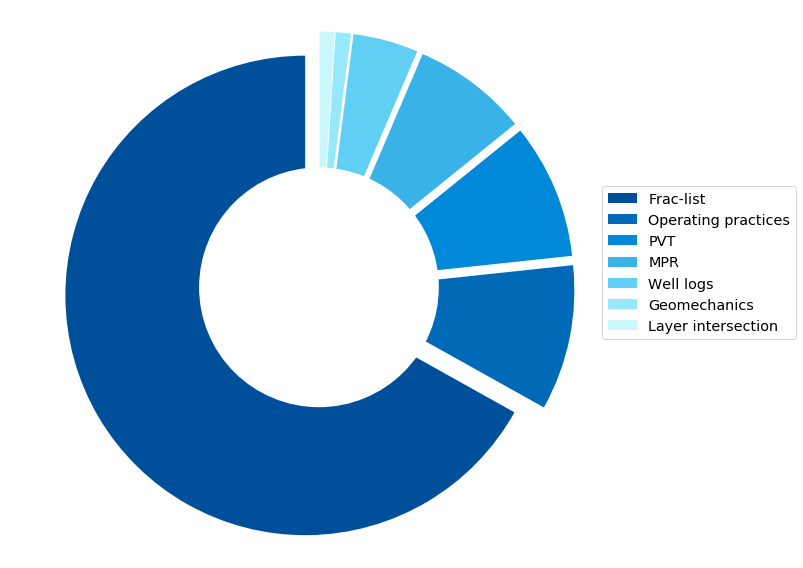}
\caption{Distribution of the initial data}
\label{distribution_base}
\end{figure}

We collect all necessary information from the following sources (Fig. \ref{distribution_base}):
\begin{itemize}
    \item Frac-list --- a document with a general description of the process and the main stages of loading;
    \item MPR (monthly production report) --- a table with production history data collected monthly after the final commissioning;
    \item Operating practices --- geological and technical data collected monthly;
    \item Geomechanics data --- principal stress anisotropy, Poisson ratio, strain modules for formations;
    \item PVT --- a general physical properties of the fluids in the reservoir;
    \item Layer intersection data;
    \item Well log interpretation data.
\end{itemize} 

Frac-list was selected as the key source of data due to the volume of crucial stage-by-stage data and existence of all ID keys, such as the field, the well, the reservoir layer and the date of HF operation. It is worth mentioning that the frac-list is full of manually filled parameters (human errors expected). Moreover, operations ended prematurely due to STOP or a screen-out are not necessarily tagged, making the problem more complex. 

\begin{figure}[h!]
\includegraphics[width=8cm]{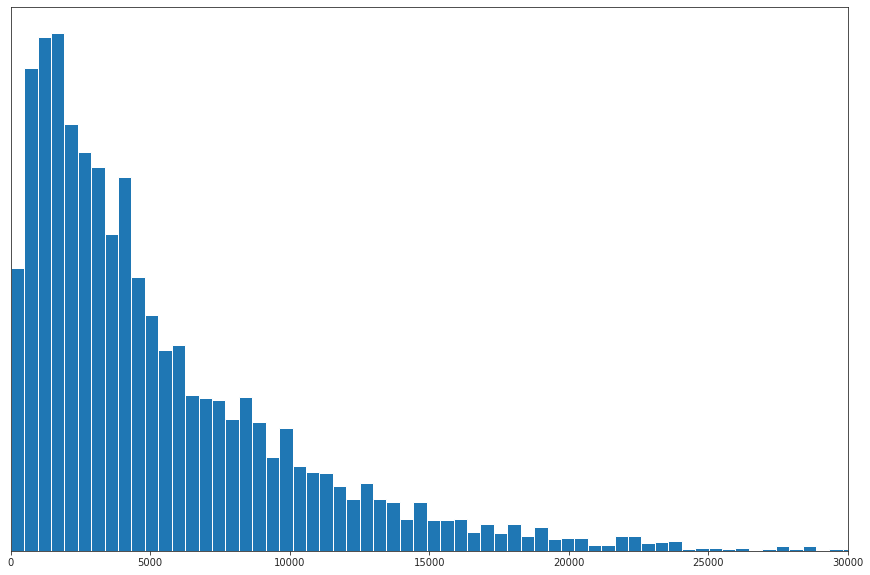}
\caption{Distribution of the 12 month production values}
\label{distribution_oil}
\end{figure}

Every source from the list above was processed individually depending on the specifics before merging them with each other. Particularly, monthly data were consolidated in 3-, 6- and 12-months slices. Fig. \ref{distribution_oil} shows distribution of cumulative oil production for 12 months (distributions for 3 and 6 months have the same form).

\begin{figure}[h!]
\includegraphics[width=8cm]{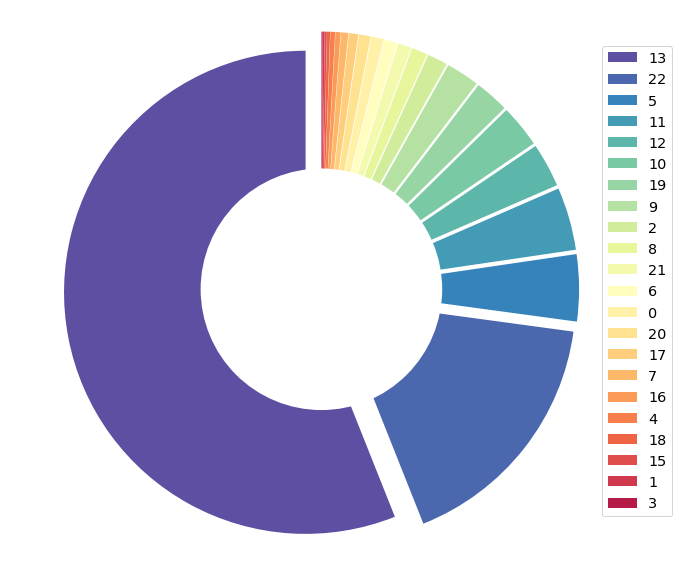}
\caption{Distribution of wells by oilfields}
\label{distribution_field}
\end{figure}

Some illustrative numbers of the initial database are presented in Table \ref{stats} and Figure \ref{distribution_field}. We show data distribution as per different oilfields, where each field is coded with a number (we avoid specific oilfield names for confidentiality reasons, in agreement with the operator). It is worth mentioning that the word \textit{operation} (in legend and tables) refers to the entire stimulation treatment, which may be a single stage fracturing on a vertical well or a multi-stage fracturing on a near-horizontal well. Then, a multi-stage treatment (operation) is divided into different stages. Each stage is characterized by the set of frac design parameters, but the transient pumping schedule within an individual stage is not (yet) considered in the present study. 

Also, the entire treatment on the layer could be repeated after a while in order to increase oil production, i.e. refracturing operation. According to data, fracturing treatment may be carried out up to five times on some wells. Thus, 36\% of total number operations has production history before the entire treatment.

\begin{table}[]
\centering
\begin{tabular}{|c|c|}
\hline
{\bf Parameter} & {\bf Numerical value }\\ \hline
Observation period & 2013 -- 2019 \\ \hline
Number of oil fields & 23 \\ \hline \hline
Number of wells & 5425 \\ \hline
-- vertical  \& directional & 4111 \\ 
-- horizontal & 1314 \\ \hline \hline
Number of fracturing operations & 6687 \\ \hline
-- single-stage treatment & 3177 \\ 
-- multi-stage treatment & 3510 \\ \hline 
-- refracturing operations (out of total) &  2431 \\ \hline 
Number of STOPs (e.g. screenout) & 797 \\ \hline \hline
Initial number of input parameters & 296 \\ \hline
Final $x$ vector of input parameters & 92 \\ \hline
-- formation  & 36 \\
-- well &  12 \\
-- frac design   & 44 \\
\hline
Number of production parameters  & 16 \\ \hline
\end{tabular}
\caption{Statistics of the database}
\label{stats}
\end{table}

\subsection{Matching database}

When merging the data from different sources, there is often a lack of a uniform template for different databases. To resolve this issue, we used regular expression algorithms and morphological analysis to identify typos. This approach allowed us to automate the process of data preparation and to make it rigorous.

To isolate typos that are inevitable in large databases, which are filled by different individuals, we created ``dictionaries'' for all sorts of categorical variables (features). With the help of the Levenshtein distance \cite{levenshtein} we found the word analogues that were considered equal. Since the ``dictionary'' we used was not very large, we applied the brute-force search strategy, which showed high efficiency. 

\begin{figure}[h!]
\centering
\includegraphics[width=9cm]{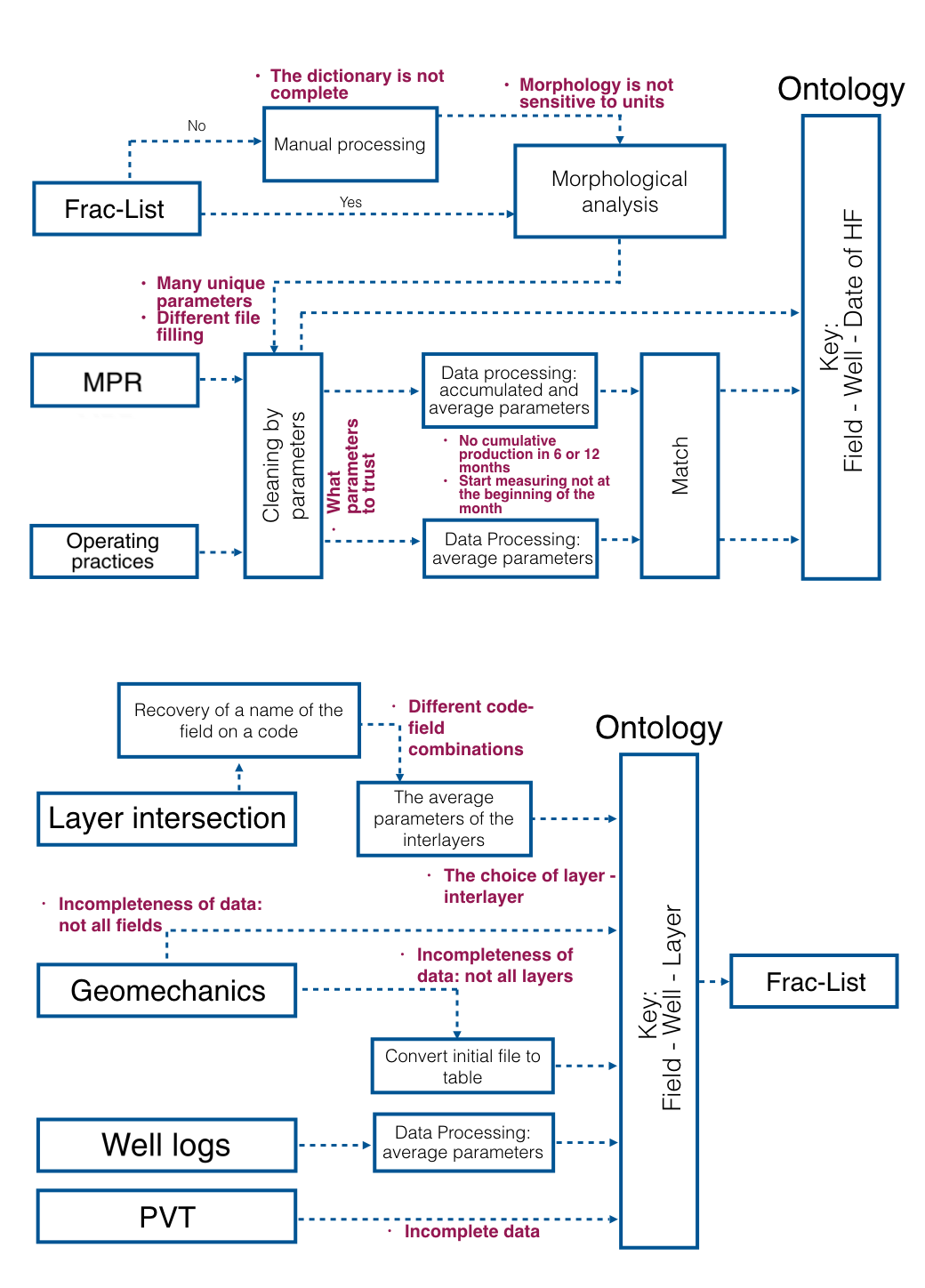}
\caption{Block diagram of the database creation}
\label{scheme}
\end{figure}

Figure \ref{scheme} shows the structure of database and its sources.

\subsection{Rounding/Averaging the values within database}

Some data sources, such as well logs, appear in a raw format, where parameters (permeability, porosity, etc.) are defined for each small interval (of the length $\approx 0.3$ m) all over the entire length of a well, where measurements were taken. Such parameters need some averaging as we need them single-valued for each well.

Averaging of these features can be handled as follows:
\begin{itemize}
    \item Porosity, permeability, clay content, oil saturation: average and mean per perforation interval and over the layer;
    \item kh: mean per perforation and layer;
    \item NTG: the total amount of pay footage divided by the total thickness of the interval for perforation and layer;
    \item Stratification factor: number of impermeable intervals per layer and perforation.
\end{itemize}

For these parameters, there is a limitation in accuracy, defined by the precision of typical well logging, which is about 30cm (hence, for determining the stratification factor one cannot detect an impermeable interval below this threshold).

As for the case of multi-stage fracturing, fluid, proppant, breaker amounts and fracture parameters (width, length and height) were summed up, and other parameters were averaged. In some cases, a well has multiple perforation intervals at different layers (multilateral wells, for example). There are production reports for every layer, distributed by its net pay. In these cases, we summarise the data from different layers (as the measurements are conducted along the entire well).

\subsection{Database cleanup}

For categorical features, we can mainly restore actual values in case of typos, whereas this is not always the case for real-valued features. 
Also very sensitive sensors (in logging) can show several values (or range of values) for a certain interval, and all of them are recorded as a characteristic for a given interval (for example, 1000-2000). However, in our database we need single-valued parameters.

As a result, we delete erroneous and uncharacteristic values: instead of magnitudes that were informational noise (e.g., a string value in a numeric parameter) a Not-a-Number value (NaN) was used. For features values, initially represented by ranges, corresponding average values were used in order to make them unambiguous and at the same time not to add extra noise to the dataset.
In order to keep our feature values in a confident range we used an expert opinion on the final compilations. For each parameter, we had a continuous exchange with a corresponding expert in field data working with the operator to crosscheck specific margins for each a zone of interest: field, formation, pad, well.

\subsection{Categorical features}

In the entire database, the number of categorical features is equal to 22. If we use one-hot encoding \cite{harris2010digital} for each unique value of each input categorical feature, the feature space is expanded with 3188 additional binary features. This leads to the curse of dimensionality problem \cite{bellman2015applied}, and, obviously, increases the calculation time and risk of overfitting. Therefore, for categorical features, which usually denote the name of the proppant for HF, we left the main name of the manufacturer and the number of the stage in which this particular proppant was used. This approach allows one to indirectly save the name and size of the proppant. Thus, the binary space dimension of categorical features increases only up to 257, which allowed us to speed up training of the ML model and to improve overall prediction accuracy.

\subsection{Handling outliers}

Outliers, i.e. objects that are very different from the most of observations in the data set, often introduce additional noise to an ML algorithm \cite{outlier2}. Outliers can be classified into three types: data errors (measurement inaccuracies, rounding, incorrect records), which occur especially often in case of field data; the presence of noise in objects descriptions; suspiciously good or bad wells; the presence of objects from other populations, e.g., corresponding to significantly different field geologies.

To effectively detect such observations we used several techniques. First of all, we used statistical methods to analyse data distribution along different dimensions and detected outliers by estimating the kurtosis measure and other statistics. 

Secondly, we used clustering. Clustering was carried out using the Density-based spatial clustering of applications with noise (DBSCAN) algorithm \cite{dbscan}, because it does not require an a priori number of clusters to be specified in advance, and is able to find clusters of arbitrary shape, even completely surrounded, but not connected. Even more importantly, DBSCAN is quite robust to outliers. As a result, outliers are concentrated in  small blobs of observations.

Another method that eliminated more than a hundred questionable values was an anomaly detection method called Isolation Forest \cite{iforest}, which is a variation of a random forest. The idea of the algorithm is that outliers fall into the leaves in the early stages of a tree branching and can be isolated from the main data sample.



\subsection{Filling missing values (NaNs)}

\begin{figure}[h!]
\includegraphics[width=8cm]{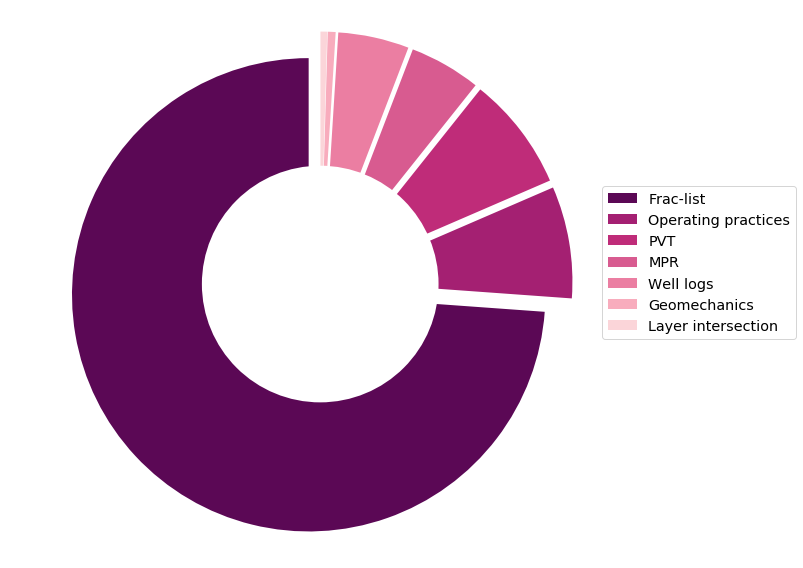}
\caption{Distribution of missing values}
\label{distribution_nans}
\end{figure}

It is very often that certain features of some objects of the field data sets are absent or corrupted. Moreover, many of ML algorithms like SVM regression or ANNs require all feature values to be known. Considering the structure of the data sources, we could expect that the frac-list has contributed to the majority of such cases of data incompleteness, since this document contains most of the useful data yet it is typically filled in manually, hence it is highly dependent on the quality of the filling process (Fig. \ref{distribution_base} \& Fig. \ref{distribution_nans}).


As a result there is a number of methods that allow one to fill in the missing values. However, it should also be noted that most approaches can be overly artificial and may not improve the final quality.

We test several approaches to fill in missing values within the framework of the regression problem under consideration:
\begin{itemize}
    \item  dropping objects containing more NaNs within an object (j-th row of a matrix representing the data set) than a certain threshold (65\%). For example, if a well has 33 missing feature values out of 50, then we drop it. Among other imputation methods described below, this would keep the database as original as possible;
    \item filling NaNs of i-th parameter by the average for the wells in a well cluster, that are grouped by geography. The reason for selecting this method of filling in is that the wells of the same cluster have similar frac designs and geology properties of the reservoir layer;
    \item filling missing values by applying imputation via collaborative filtering (CF) \cite{colab}. CF is a technique used by recommender systems, which makes prediction of absent values with the use of mathematical methods. According to our research, the best results were shown by non-negative matrix factorization (NNMF) and truncated singular value decomposition (TSVD). Worth noting, NNMF cannot handle negative values, such as skin-factor.
    \item applying unsupervised learning to define similarity clusters and filling NaNs of i-th parameter by the mean of the cluster. In other words, the average of the feature is taken not from the entire database, but from the cluster, which allows us to estimate the missing value more accurately.
\end{itemize}

\section{ML algorithms implemented in the present work}
\subsection{Regression}
Once the database is created, four imputation methods of filling NaNs are evaluated in terms of their performance. After applying these imputation methods to the database, the results of the best tuned ML algorithms for regression problem are as follows ($R^2$):
\begin{itemize}
\item Matrix factorization: $R^2=0.773$; 
\item Dropping the entire row, if NaN's count more than 65\% in that row: $R^2=0.688$;
\item Filling with mean values of the well pad: $R^2=0.607$;
\item Filling with mean values of the cluster: $R^2=0.577$.
\end{itemize}
Comparison between different ML algorithms with these filling methods can be seen in Figs.~\ref{FigPDrop}, \ref{FigPPad}, \ref{FigMPad} and \ref{FigPClusters}.
Matrix factorization appeared to be the most effective method, so it was applied to the entire dataset. Handling negative values (skin-factor) has been done via introducing a binary parameter, which shows whether the skin is negative or not.

Then, several ML models were chosen to predict cumulative  oil production over 3 months. The target function distribution is shown in Fig.~\ref{distribution_oil}.


The following ML regression algorithms were used: SVM, KNN, ANN, Decision Trees, and various types of ensembles based on decision trees such as Random Forest, ExtraTrees, CatBoost, LGBM and XGBoost.

Each model was trained on a subsample with cross validation on 5 folds. Then, models were tested on a separate (hold-out) sample. All these sets were shuffled and had similar target value distributions. Most of the ML models are decision tree-based and, hence, have important advantages: they are fast, able to work with any number of objects and features (including categorical ones), can process data sets with NaNs and have a small number of hyperparameters.

Each experiment is conducted two times on four data sets constructed using  different imputation techniques: 
\begin{itemize}
    \item on the entire dataset containing information about 5425 wells. Here, we used hyperparameters of the regression algorithms set to their default values first. Then, after figuring out the best imputation technique, we proceed to the next experimental setup described below;
    \item on wells from one field only; again, we used default hyperparameters of the regression algorithms.
\end{itemize}

The reason to use two experimental setups is to check if more homogeneous data set enhances predictive performance of the model.

Then, we take the best performing methods based on the $R^2$ on test set of each experiment, tune their hyperparameters via the grid search, and combine them into an ensemble to further improve the results. If the result of the ensemble of models is worse than the single best regressor, then we are taking the results of the best regressor.

\subsection{Feature analysis}
Feature importance analysis is performed for an ensemble of the best algorithms. OVAT analysis is carried out to see how the target varies with the variation of the design parameters. In addition, if the feature rankings of both methods are more or less similar, then we may proceed to parameter reduction. With the available feature importance values, we iterate over a range to remove less important parameters and then calculate the $R^2$ score. This procedure is important for the design optimization (which would be considered in the second psrt of this research in another article), because it reduces the dimensionality of the problem while keeping the best score.

\subsection{Uncertainty Quantification}
Finally, uncertainty quantification is done for the model metric (the determination coefficient $R^2$) by running the model multiple times for different bootstrapped samples. We can represent a result of this step in the following form:  95\% probability chance that the value of $R^2$ is located within the given interval. The scheme of the forward problem methodology is depicted on Fig.~\ref{FigForward} (see Appendix).

\section{Results and Discussion}

\subsection{Filling missing values and clustering}

By applying the first method of missing data imputation, we droppped the rows with more than 42 NaNs. Then, the rest of the NaNs for other objects in the data set are filled with their mean values. 

\begin{figure}[h!]
\includegraphics[width=9cm]{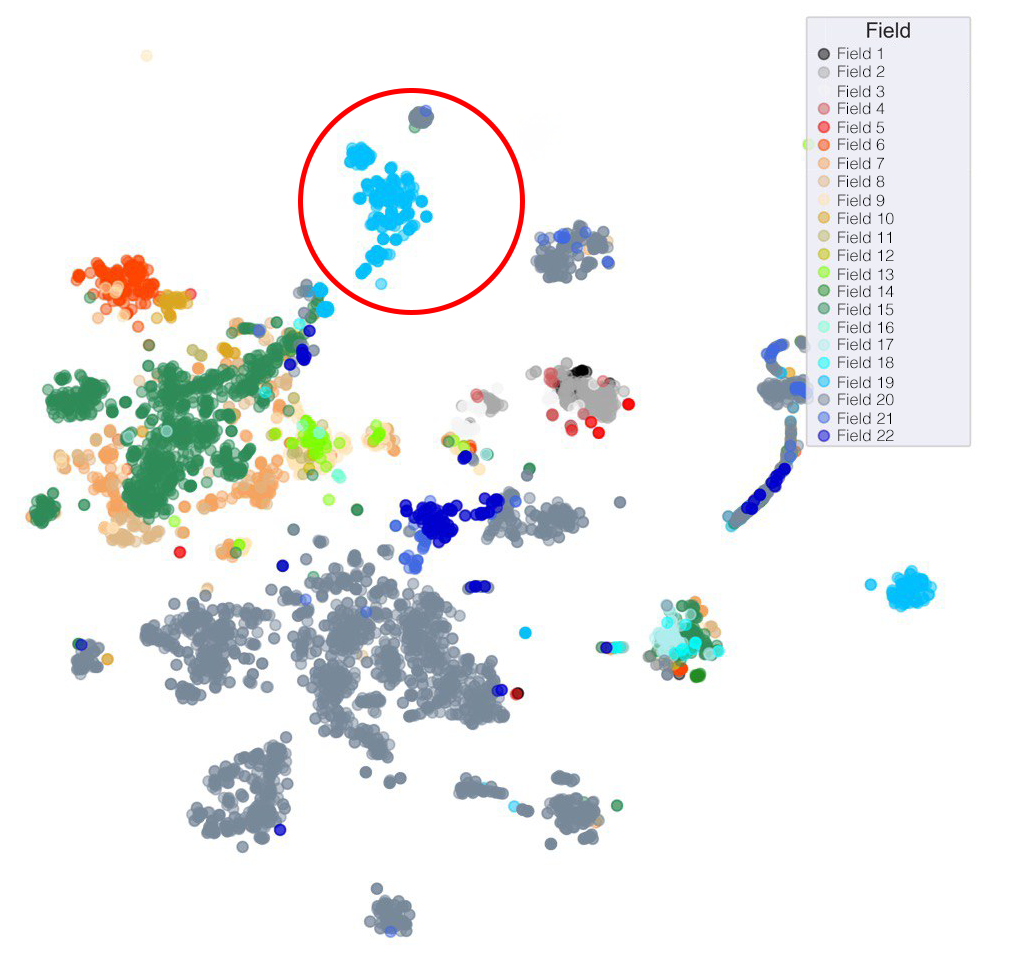}
\caption{t-SNE visualization plot of the whole database for imputing NaNs by clusters}
\label{FigClustersWhole}
\end{figure}

Fig.~\ref{FigClustersWhole} shows how the entire database is clustered with DBSCAN algorithm. Since the algorithm itself cannot handle missing values, we recover them using the collaborative filtering. Then we assign cluster labels for each well to the original data set. To visualize clusters, t-SNE is applied to transform data space into 2D and build a scatter plot. As seen from the figure, there are 3 groups in total with the biggest cluster marked as ``2''


\subsection{Regression}


The results of the four imputation methods on the entire database are shown in Appendix. The $R^2$ is calculated for the test sample for 9 regression algorithms.

We can see that:
\begin{itemize}
    \item The family of decision-tree based algorithms show better accuracy than other approaches. CatBoost algorithm (based on gradient boosted decision trees) outperforms all other methods;
    \item Some of the ML algorithms like SVM and ANN  resulted in negative $R^2$, which is interpreted as poor prediction accuracy. The possible explanation is that both methods are preferred when there are homogeneous/hierarchical features like images, text, or audio, which is not our case;
    \item The best imputation technique is collaborative filtering;
    \item Based on the log scale regression plot, a relatively large amount of errors comes from the points with too low or too high oil production rates. The possible solution of the problem is to perform regression for different clusters;
    \item Once the imputed dataset was split into 3 clusters, we found out that cluster ``0'' and ``3'' contain a relatively small amount of samples ($\sim 800$ wells). Therefore, it has become evident that if we construct a separate predictive model on each cluster this can result in overfitting even with a simple model (KNN). Thus, the test $R^2$ score appears to be 0.750 for both clusters. For the largest cluster (number 2), which contains 1844 objects, the $R^2$ test score reached values higher than 0.850, and the overall $R^2$ is 0.8, which is less than for the model trained on the whole dataset.
\end{itemize}

\subsection{Hyperparameter search and ensemble of models}

Since CatBoost performs better than other boosting algorithms and itself is an ensemble model, there is no need to create ensemble of models. What needs to be done is to tune the model to achieve higher score. Since our model is overfitting, it is necessary to regularize it. With the help of grid search, we iterate over values of the L2 regularization of the CatBoost ranging from 0 to 2 with a 0.1 step. The tree depth is ranging from 2 to 16. These hyperparameters have the highest impact on the overall test $R^2$.

\begin{figure}[h!]
\includegraphics[width=9.2cm]{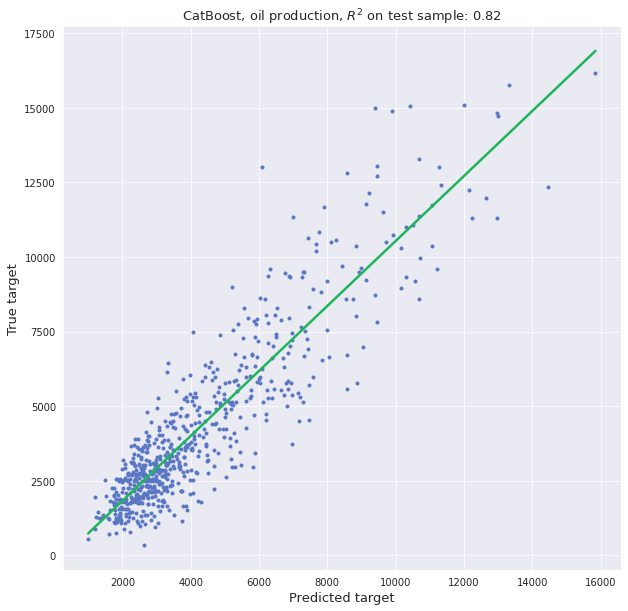}
\caption{Regression plot for the best model on test set}
\label{FigMatrixBest}
\end{figure}

\begin{figure}[h!]
\includegraphics[width=9.2cm]{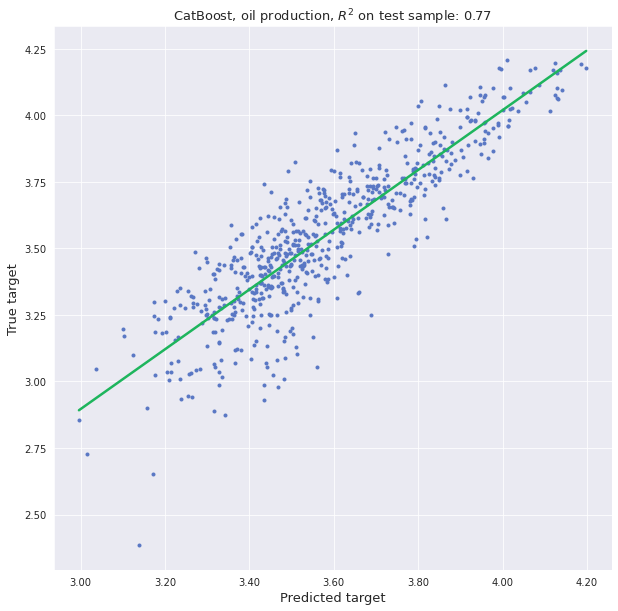}
\caption{Logarithmic regression plot for the best model on test set}
\label{FigMatrixBestLog}
\end{figure}

Below is the list of optimal hyperparameters, for which the final accuracy is 0.815 on the test set for target (Fig.~\ref{FigMatrixBest}) and 0.771 (after exponentiating) for target logarithm (Fig.~\ref{FigMatrixBestLog}):
\begin{itemize}
    \item depth = 7;
    \item l2 leaf reg = 0.6;
    \item learning rate = 0.02;
    \item od type = 'Iter' --  the overfitting detector, the model stops the training earlier than the training parameters enact;
    \item od wait = 5 -- the number of iterations to continue the training after the iteration with the optimal metric value.
\end{itemize}

\subsection{Feature analysis}

\begin{figure}[h!]
\centering
\includegraphics[width=9cm]{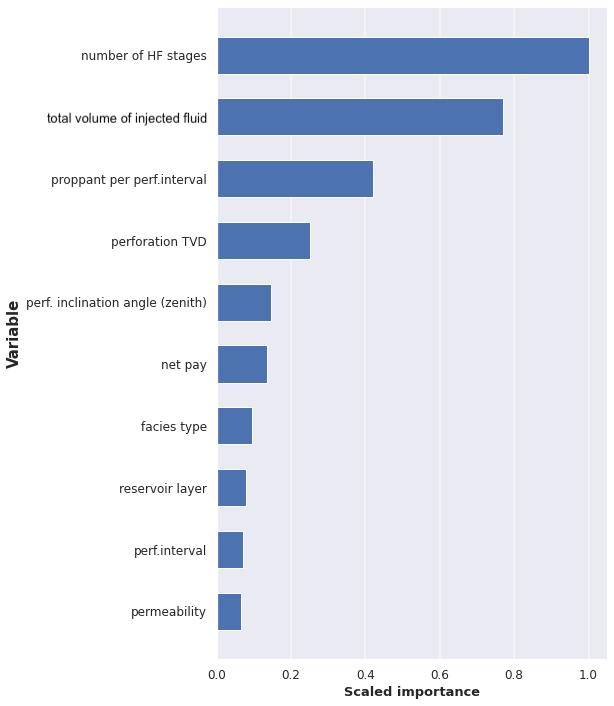}
\caption{Feature importance}
\label{FigMatrixFeatImp}
\centering
\end{figure}

Figure~\ref{FigMatrixFeatImp} shows feature importance for the 25 important features. The top five important overall features are: 
\begin{enumerate}
    \item number of stages in the multi-stage hydraulic fracturing treatment;
    \item proppant mass per meter of perforated interval;
    \item volume of the injected fluid;
    \item net pay; 
    \item perforation true vertical depth.
\end{enumerate}

\begin{figure}[h!]
\includegraphics[width=9cm]{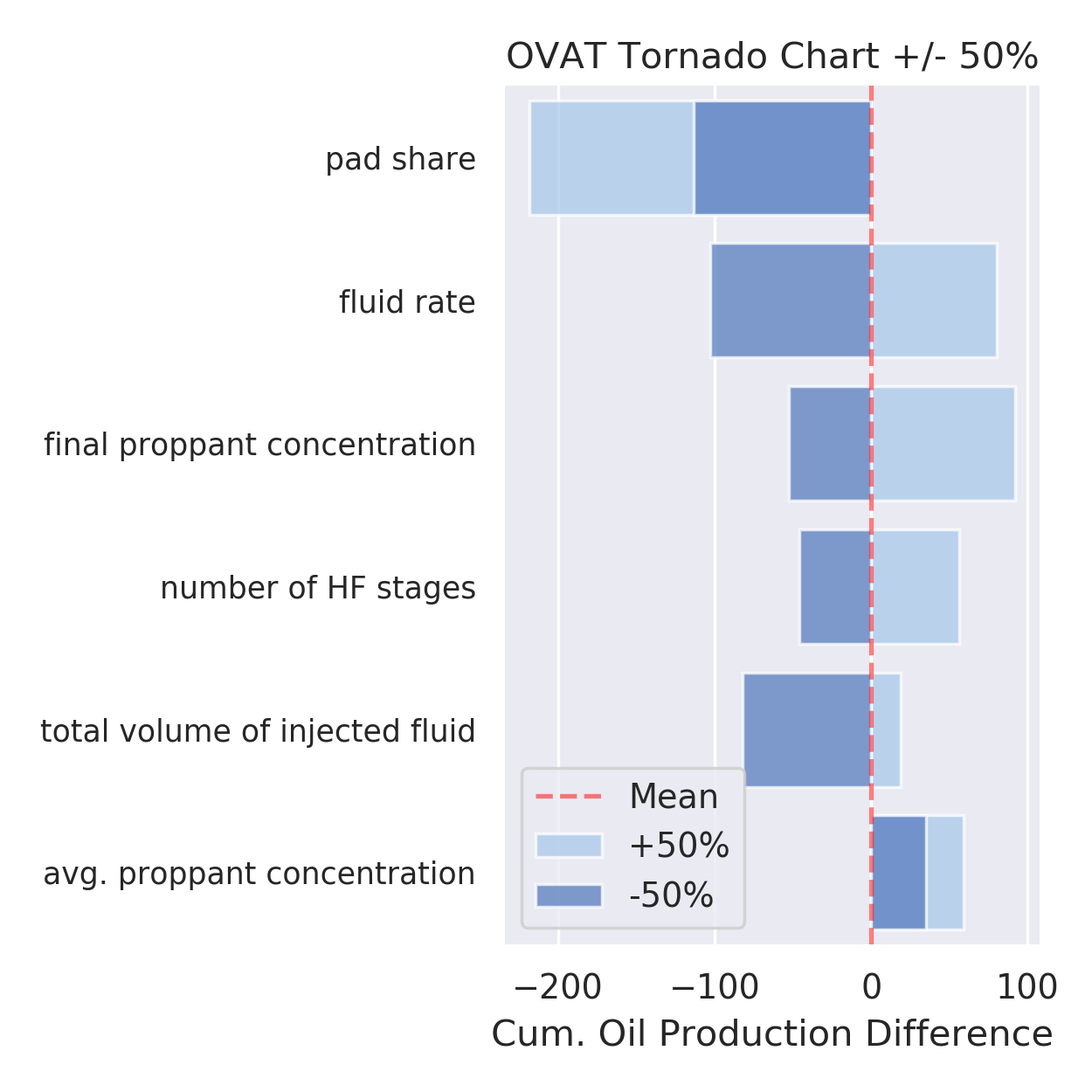}
\caption{Tornado chart of the OVAT analysis}
\label{FigMatrixTornado}
\end{figure}

Figure~\ref{FigMatrixTornado} shows a Tornado chart of the OVAT. The most relevant features are indicated as bars, and the red dashed line is a target where all parameters are taken at their mean values over the data set. To interpret the graph, consider the top feature, pad share (which is the ratio of the pad volume to the volume of injected fluid). The dark blue means the difference between the target with the ``average parameter value'' and the target with the pad share parameter decreased by 50\%, while the rest of the parameters are kept at their average values.

The feature importance analysis within the Catboost model is carried out for the entire feature list, while the OVAT analysis is conducted for the design features only. The reason for performing the OVAT on the design features only is due to the problem objective, where our goal is to vary design parameters to maximize the target. Moreover, the design features deviate from their means, which is oftentimes not true for different types of wells. In other words, the limitation of OVAT is that we have to deviate the {\it i}-th parameter from its average, while some wells have the value of the {\it i}-th feature far from the mean. Hence, the OVAT is more applicable for the design optimization problem and is not consistent with the feature importance analysis. An example of inconsistency is the number of HF stages, where its ranking within the feature importance (Fig~\ref{FigMatrixFeatImp}) is 1 while its ranking on the tornado chart (Fig~\ref{FigMatrixTornado}) is 4 among other design parameters. To summarize, OVAT is not representative for verifying feature importance, while it is suitable for getting target value sensitivity to the variation of a single parameter on a particular HF operation (with all other features fixed at their mean values).

\subsection{Parameter selection}

\begin{figure}[h!]
\centering
\includegraphics[width=10cm]{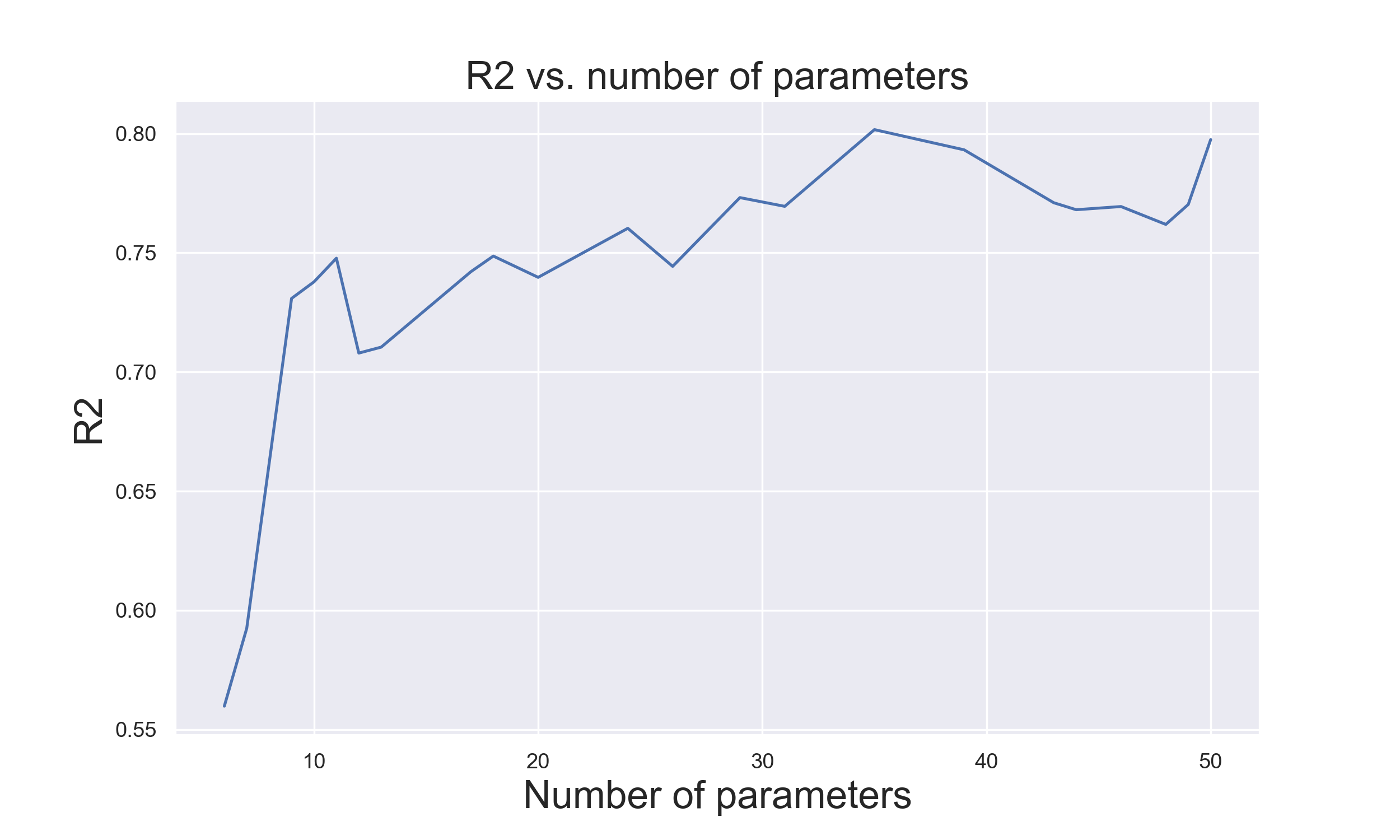}
\caption{Score vs. number of important parameters}
\label{FigR2vsParams}
\end{figure}

The dependence of the predictive capability of the model on the number of parameters taken into account is obtained based on the feature importance analysis described above (see Fig ~\ref{FigR2vsParams} for illustration). Based on the results, the dimensionality of the problem is reduced from 50 to 35 parameters, which makes the design optimization problem more tractable. In addition, the $R^2$-score does not improve, when the input vector dimension increases above 35 parameters. As one can see, not all design features presented on OVAT tornado chart are in the top-10 important features. However, design parameters will be still added to the input data since they are going to be optimized, which is the main goal of the second part of this work. 

\subsection{Uncertainty Quantification}

\begin{figure}[h!]
\centering
\includegraphics[width=10cm]{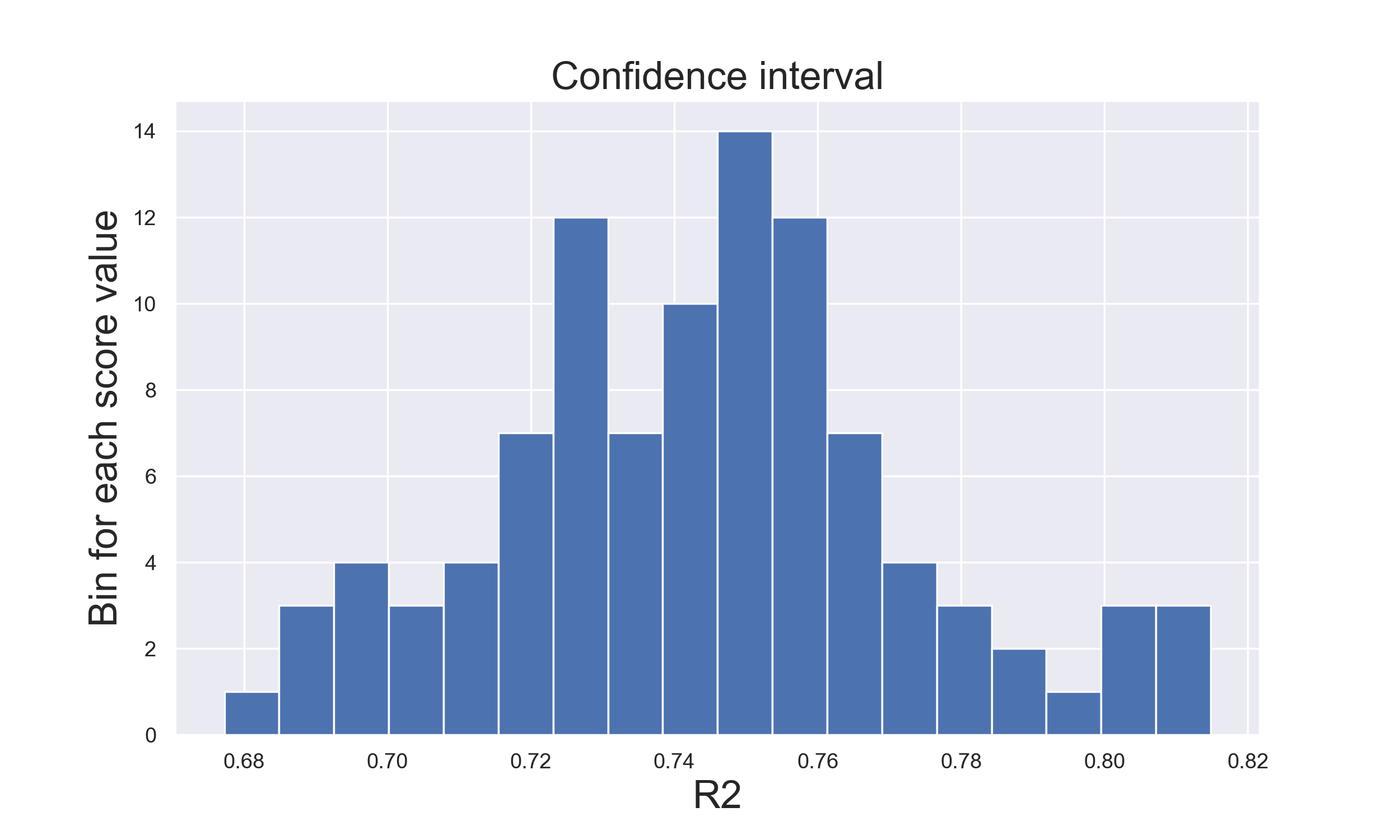}
\caption{Confidence interval}
\label{FigConfidenceInterval}
\end{figure}

To completely describe the results, uncertainty has to be determined via a confidence interval. To calculate the confidence interval for the ensemble model $R^2$, the bootstrap is used for 100 iterations, where each subset is 75\% the size of the entire database. As the result, the calculated confidence interval shows that there is a 95\% likelihood that the confidence interval 0.680 and 0.815 covers the true $R^2$ of the model (Fig ~\ref{FigConfidenceInterval}). As it can be seen, the most frequently encountered result lies between 0.74 and 0.75. Therefore, the overall $R^2$ has to be increased or the variance of the $R^2$ confidence interval has to be decreased to improve the results.

This paper anticipated the further study of an inverse problem, where we will consider an optimization problem on the set of HF design parameters with maximized cumulative production and also we will identify a proper metric, e.g. Q/CAPEX or NPV, to be able to distinguish between economically justified scenarios and high production just due to massive (and expensive) proppant pumping.

\subsection{Comparison with other studies and discussion}
Here we would like to discuss our results in light of the recent study \cite{wang2019insights}, which appears to be most relevant, to the best of our knowledge in open literature. First, we were focused on a self-consistent, comprehensive database by designing its architecture with ML methods in mind. Second, the forward problem is proceeded with different heuristic imputation methods, and the best one is further applied together with an ML prediction method. We carefully performed various cross-validation and test accuracy estimates to minimize the effects of overfitting in our model, and get higher $R^2$ of 0.815. Speaking about the ML feature importance, we implemented the OVAT analysis, and the results of the method showed the ranking and the correlation of design parameters and the target (Fig ~\ref{FigMatrixTornado}), where the most sensible feature occurred to be a pad share with a negative correlation with the target. Ensuring the robustness of the model, we performed a parameter selection similar to the recursive feature elimination with cross validation (RFECV) mentioned in \cite{wang2019insights}, that reduced our feature space from the initial database 50 parameters (selected manually) to only 35. In addition, uncertainty quantification is performed to fully describe our accuracy score and showed model's limitations.

Even though the two studies have much in common, the key difference comes from the database architecture. We built the database from scratch focusing on its applicability for the ML. Such approach not only resulted in better forward problem results, but also gives an opportunity to consider different combinations of input parameters. By further investigating and engineering the most appropriate features of HF design, we should be able to obtain a full description of the well with various types of parameters (PVT, well logs, geomechanics, MPR and others). Better accuracy would open the way to more accurate inverse problem results, which will be discussed in the next part of the work.

The study~\cite{2019RN-selection} considered an oilfield with about 1500 wells. The database after pre-processing included 477 frac jobs   (compare with 6687 in our case). All these jobs were conducted on the wells, which have already been producing (refrac treatments), so its production before treatment was known. This makes the production forecast problem much easier to solve, when the level of production prior to fracturing is an input parameter. Nevertheless, for comparison, we have 2431 of refracturing operations in our database, and our model yields MAPE (Mean Absolute Percentage Error) of less than 10\%  and $R^2>0.85$ on refracs, compared to 23\% MAPE in~\cite{2019RN-selection}. In addition, \cite{2019RN-selection} reported that the production rate prior to fracturing is on the top of the feature importance list for predictive ability of the model.

\section{Conclusions}

We presented a global framework for constructing ML algorithms on digital field database for the purposes of HF design optimization. This is Part I of the study, which is specifically aimed at data gathering, cleaning, systematization and pre-processing with the purpose of forward problem solution (production forecast based on fracture design parameters). We discussed in detail the issues that raise on the way towards constructing a digital field database, which integrated three major parts coming from essentially different sources: reservoir geology, HF design and production data. The resulting database covers more than 5000 oil wells in Western Siberia (drilled in the period 2013-2019), including vertical, directional and horizontal wells, stimulated with fracturing treatments. This is a remarkably representative dataset, compared to the majority of open literature on the subject. The overall $x$-vector characterizing a well (data point) contains 92 parameters, including 36 parameters for formation, 12 for the wellbore, and 44 for the fracturing design. The input vector is reduced to 35 parameters after recursive feature importance analysis and elimination. Production is characterized by 16 different parameters. 

The forward problem of predicting the production rate based on fracturing design parameters is solved using the most widely used ML algorithms. Cross-comparison revealed that decision tree based models outperformed the others due to high heterogeneity of the input parameters. As the result of solving the forward problem, the accuracy of predicting cumulative oil production is $R^2$=0.815 achieved by the CatBoost algorithm. Regarding relative feature importance within the model, the top ten important parameters are:
\begin{itemize}
\item number of stages in a multistage treatment;
\item  volume of injected fluid;
\item proppant mass per meter of perforated interval;
\item perforation true vertical depth;
\item perforation zenith angle;
\item reservoir net pay;
\item geological facies;
\item reservoir layer;
\item perforated interval;
\item formation permeability.
\end{itemize}

From the OVAT analysis, it follows that the following possible patterns in the average multi-stage HF treatment can be identified (with all parameters at its mean values):
\begin{itemize}
    \item Mean value of the pad share is optimal for an average treatment. Deviations from this value have negative effect on production;
    
    \item Increasing the fluid rate increases production and vice versa;
    
    \item Mean final proppant concentration possibly was selected below optimum, comparing to the optimum value, which is less than average. In addition, average proppant concentration is probably systematically selected below optimum value, too (by frac design engineers planning the treatment);
    
    \item Building conclusions on the results of both OVAT and feature importance analysis, we come to the conjecture that the volume of injected fluid is one of the most important features.
\end{itemize}

In future research, this analysis will examine various mean values (comparing single-stage / multistage treatments, multilayer mutilaterals / single layer multistage, vertical / horizontal wells, etc.)

Cleaning the data and handling the missing data on real field data set appeared to be one of the most important tasks due to huge amount of errors and missing records in the original raw data. Missing values imputation via collaborative filtering technique (NNMF) allowed us to improve predictability of a model by $\sim17\%$ ($R^2$).
The highest predictive capability of the model proved to be based on a data base of wells with re-fracturing (where production before treatment is known).

The following important points need to be emphasized:
\begin{itemize}
    \item ML model completely depends on input data (data completeness, data quality, and preprocessing);

    \item Collection of field data is the most important step for the ML project aimed at an optimization of a stimulation treatment. A database, which has been properly validated, filled and verified with subject matter experts, allows one to build high-quality predictive models and make well-informed decisions, based on all the advantages of modern ML techniques;
    
    \item Data pre-processing and use of complex tuned ML models allow to achieve high accuracy. However, the results of our study on small train sets versus full data set show that this accuracy does not always indicate the model capability to generalize the results obtained. High accuracy reported in the literature on relatively small data is typically the consequence of overfitting. Our results were validated on a proper hold-out set, which was not used in training of the model;
    
    \item The test accuracy of the model highly depends on the number of samples and on the complexity of the ML algorithm. If either of these two does not fit together, it would lead to overfitting or underfitting. In our case, the best available ML practices are selected and tuned. Overall, the more data the better to further increase the accuracy since our best performing model is a greedy decision tree based algorithm.
\end{itemize}

Thus, an accurately formed, verified and validated field database on stimulation treatments may lead to the results that are not "ideal" (in terms of the determination coefficient), because of its inherent heterogeneities/ambiguities. 

Speaking about the forward problem in determining the cumulative production, we integrated the most novel approaches available in ML nowadays by applying clustering, model ensembles and tuning, feature importance, and uncertainty quantification.

In Part II of this work (to be reported in a separate paper) we will focus on the inverse problem of optimizing HF design parameters to maximize oil production, using the database built so far and the forward modeling algorithms developed here to predict production based on the frac design parameters. To solve the inverse problem, several approaches will be proposed: brute force optimization via GridSearch, Bayesian optimization, a heuristic application of reinforcement learning, and other non-gradient optimization methods. Cost efficiency will also be considered to find a balance between investment and return, considering the outcome of an HF job in the perspective of maximum Q/CAPEX.

\section*{Abbreviations and key terms}
ML -- Machine Learning

HF -- Hydraulic Fracturing

ANN -- Artificial Neural Network

PVT -- Pressure Volume Temperature

SVM -- Support Vector Machine

NPV -- Net Present Value

KGD -- Khristianovic-Geertsma-de Klerk

PKN -- Perkins-Kern-Nordgren

P3D -- Pseudo-three-Dimensional

NN -- Neural Network

TOC -- Total Organic Carbon

t-SNE --  t-distributed Stochastic Neighbor Embedding

OVAT -- One-Variable-At-a-Time

MPR -- Monthly Production Report

RIGIS -- well log interpretation results

STOP -- case of interrupted HF operation (e.g. due to screen-out)

NaN -- Not-a-Number value

DBSCAN --  Density-Based Spatial Clustering of Applications with Noise

CF -- Collaborative Filtering

NNMF -- Non-Negative Matrix Factorization

TSVD -- Truncated Singular Value Decomposition

KNN -- k-Nearest Neighbours

LGBM -- Light Gradient Boosting Machine

XGBoost -- eXtreme Gradient Boosting

MD -- Measured Depth

TVD -- True Vertical Depth

DFIT -- Diagnostic Fracture Injection Test

ISIP -- Initial Shut-In Pressure

NTG -- Net-To-Gross formation ratio

BH -- Bottom-Hole

CAPEX -- CAPital EXpenditure

Pad -- Viscous fracturing fluid without proppant pumped into a well before slurry to open and propagate the fracture.

Pad share -- Ratio of the pad volume to the entire volume of injected fluid.

\section*{Acknowledgements}
{The authors are grateful to the management of LLC ``Gazpromneft-STC'' for organizational and financial support of this work. The authors are particularly grateful to A.G.~Kan, M.M.~Khasanov, A.A.~Pustovskikh, M.F.~Staritsyn, I.G.~Fayzullin and A.S.~Margarit for organizational support of this initiative. The help from E.~Davidyuk and V.~Mikova in data gathering is gratefully appreciated. Insightful discussions with production stimulation engineers of the operator LLC "Gazpromneft-Khantos", in particular with R.P.~Uchuev, I.S.~Vikhman and N.~Chebykin, are appreciated.

We would like to state it explicitly that the models presented in this work are solely based on the field data provided by JSC Gazprom neft and we are grateful for the permission to publish.

Startup funds of Skolkovo Institute of Science and Technology are gratefully acknowledged by Prof. A.A.~Osiptsov.
}

\bibliographystyle{elsarticle-num}
\bibliography{MLfracturing}

\onecolumn
\newpage
\centering
\section*{Appendix}
\begin{table}[!h]
\centering
\begin{tabular}{|l|l|l|}
\hline
\multicolumn{3}{|c|}{Formation parameters}  \\ \hline
Layer  & Porosity average per perforation & Median clay content per perforation \\ \hline
Net pay  & Porosity average per layer & Median clay content per layer \\ \hline
Facies type  & Porosity median per perforation & Oil saturation average per perforation \\ \hline
Formation thickness  & Porosity median per layer & Oil saturation average per layer \\ \hline
Formation pressure  & Permeability average per perforation & Oil saturation median per perforation \\ \hline
Bubble point pressure  & Permeability average per layer & Oil saturation median per layer \\ \hline
Oil formation volume factor  & Permeability median per perforation & NTG per perforation \\ \hline
Permeability from well flow test & Permeability median per layer & NTG per layer \\ \hline
Oil viscosity & kh median per perforation & Stratification factor per perforation \\ \hline
Water viscosity & kh median per layer & Stratification factor per layer \\ \hline
Oil density & Average clay content per perforation & Formation temperature \\ \hline
Formation pressure & Average clay content per layer & Well intersection data \\ \hline
\multicolumn{3}{|c|}{Well structure}  \\ \hline
Perforation depth (MD) & Inclination angle  & Tubing diameter \\ \hline
Perforation depth (TVD)  & Well's drift direction  & Perforation density \\ \hline
Perforation interval  & Skin before/after HF  & Perforation type \\ \hline
Drainage radius  & Dimensionless productivity index (Jd)  & Inclination angle from well-logs \\ \hline
\multicolumn{3}{|c|}{HF design parameters}  \\ \hline
Number of HF stages  & Fracture permeability  & Proppant per gross height \\ \hline
Multifrac stage  & Closure gradient  & Shut-in pressure \\ \hline
Polymer type  & ISIP for displacement  & Breaker \#1 amount \\ \hline
Polymer concentration  & ISIP on DFIT & Breaker \#2 amount \\ \hline
Crosslinker type  & ISIP on main work & Breaker \#3 amount \\ \hline
Crosslinker concentration  & Delta ISIP & Pad volume \\ \hline
Polymer concentration for pad  & Effective pressure on DFIT & Fracture length \\ \hline
Fluid type for main work  & Effective pressure on main work & Fracture height \\ \hline
Breaker type \#1  & Fluid efficiency & Fracture width \\ \hline
Breaker type \#2  & Proppant concentration & Mass of proppant type \#1 \\ \hline
Breaker type \#3  & Pressure loss on friction & Mass of proppant type \#2 \\ \hline
Average pressure on main work  & Pressure loss in BH area & Mass of proppant type \#3 \\ \hline
Dimensionless fracture conductivity  & Proppant per oil-saturated height & Mass of proppant type \#4 \\ \hline
Fracture conductivity  & Proppant per effective height & Mass of proppant type \#5 \\ \hline
Mass of all proppant & Fluid volume &  \\ \hline
\multicolumn{3}{|c|}{Production data}  \\ \hline
Bottom-hole pressure  & Watercut before HF & \textbf{Cumulative oil production} (target) \\ \hline
Productivity index  & Suspended solids concentration & Cumulative fluid production \\ \hline
Dimensionless productivity index (Jd)  & Fluid rate after HF & Cumulative gas production \\ \hline
Fluid rate  & Oil rate after HF & Watercut average during production \\ \hline
Gas rate before & Watercut after HF & Operational hours during production \\ \hline
Fluid rate before HF &   &  \\ \hline
\end{tabular}

\textit{*features averages and medians per layer and perforation sourced from well log interpretation data}

\textit{**these are all parameters in the data base before feature selection}
\caption{Features used to describe a well}
\label{tabtab}
\end{table}

\begin{figure*}[h]
\centering
\includegraphics[width=12cm]{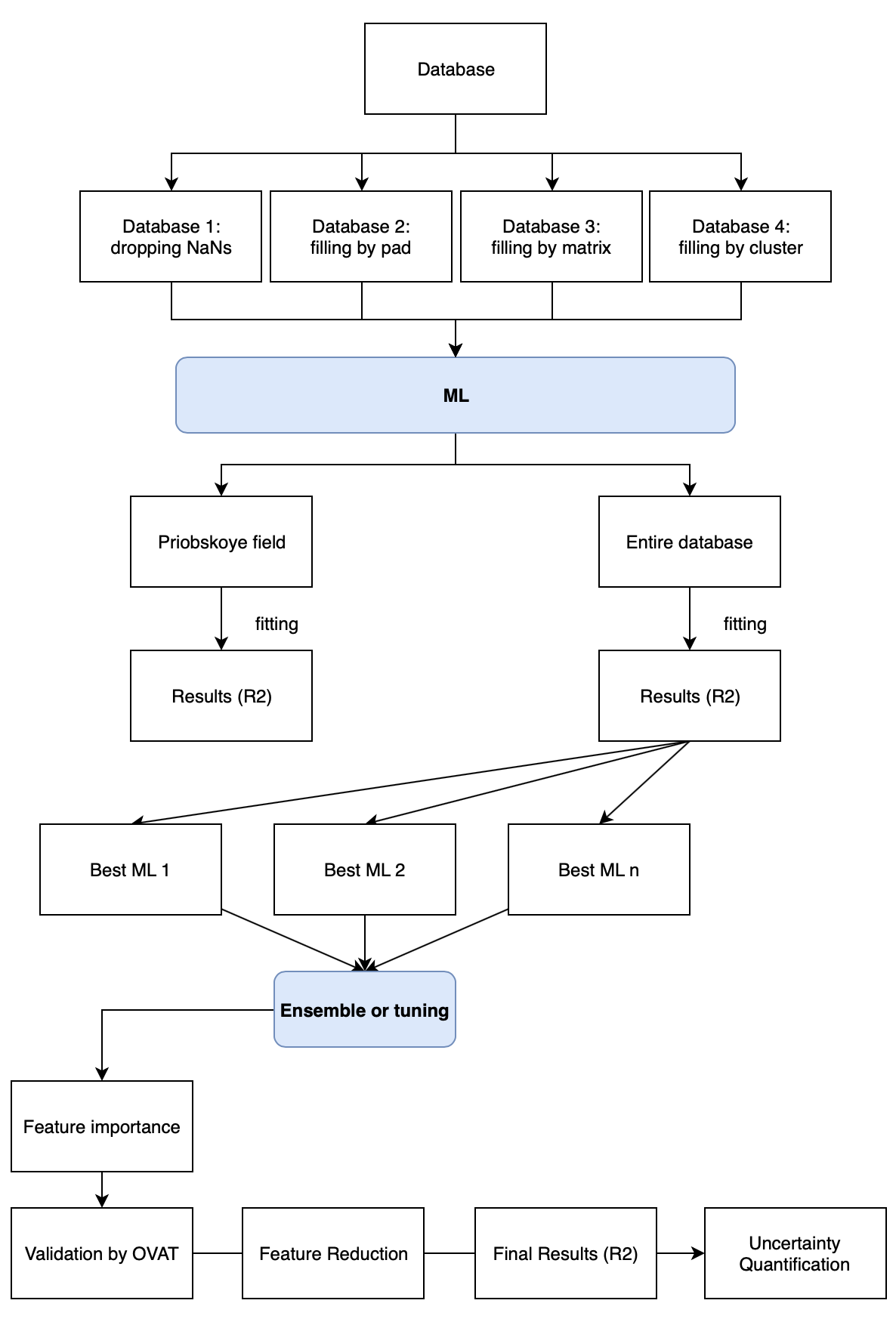}
\caption{Forward model algorithm}
\label{FigForward}
\end{figure*}

\begin{figure*}[t]
\centering
\includegraphics[width=16cm]{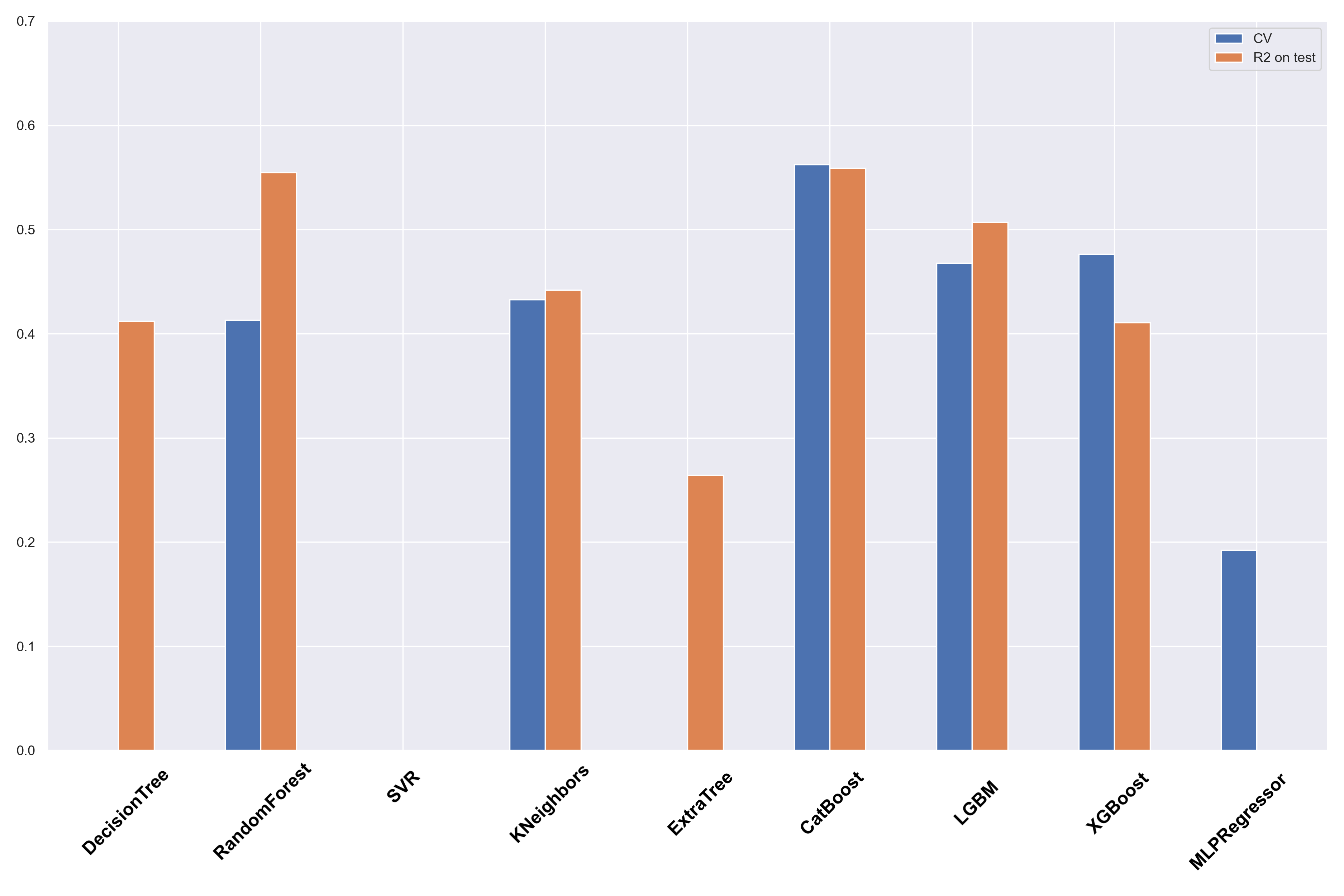}
\caption{Algorithms' performance on a test set (untuned): dropping NaNs method}
\label{FigPDrop}
\end{figure*}

\begin{figure*}[b]
\centering
\includegraphics[width=16cm]{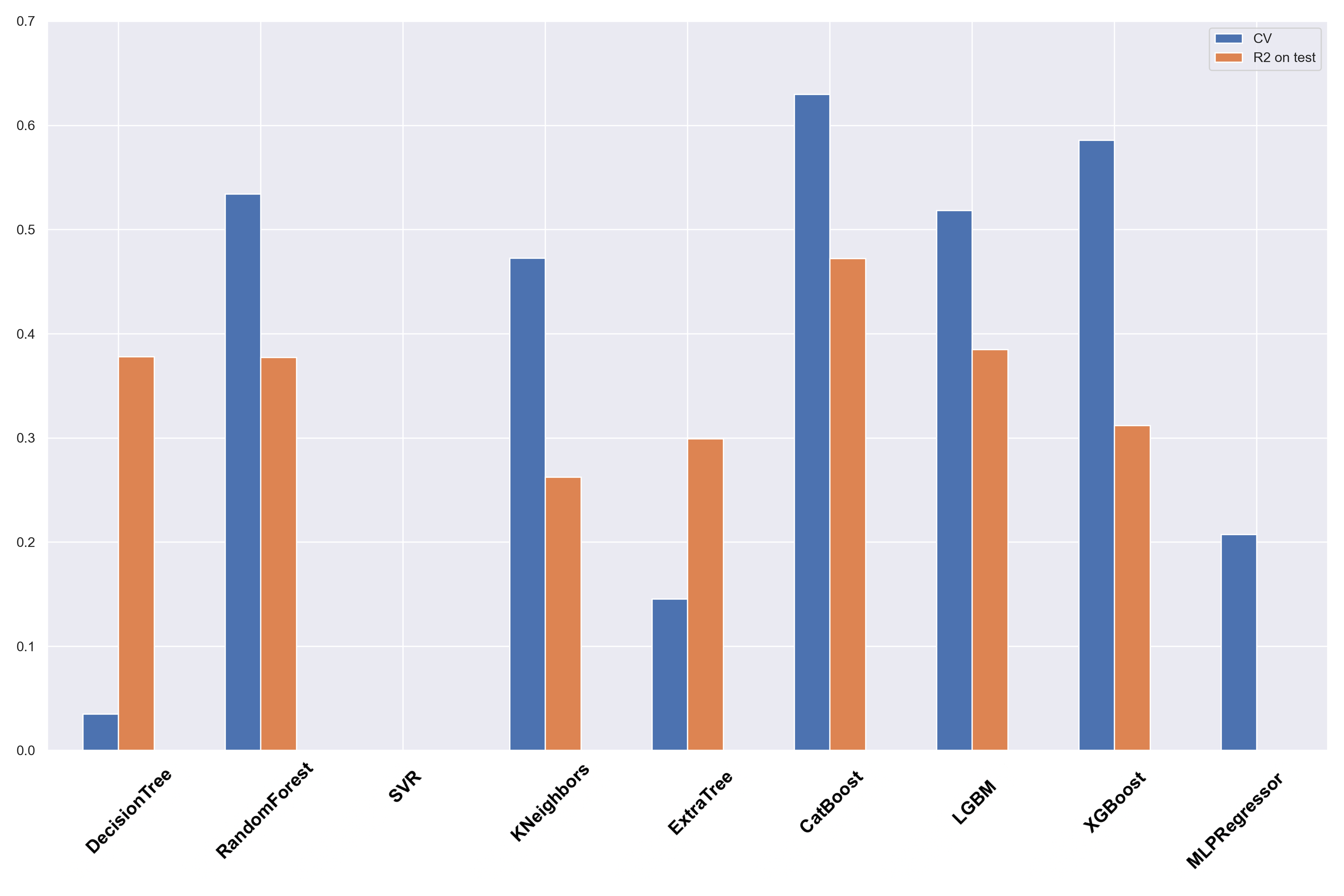}
\caption{Algorithms' performance on a test set (untuned): filling by well pad}
\label{FigPPad}
\end{figure*}

\begin{figure*}[t]
\centering
\includegraphics[width=16cm]{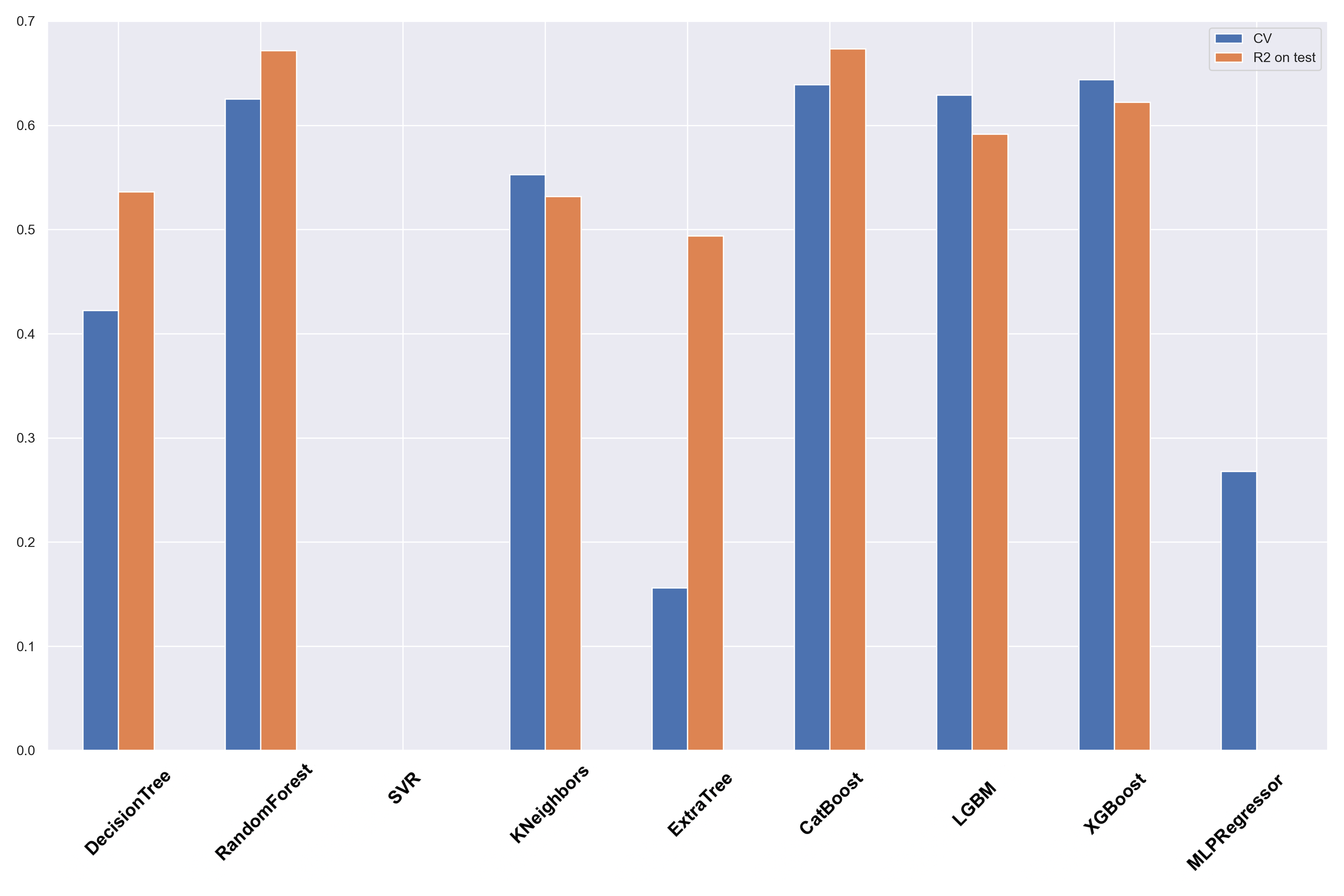}
\caption{Algorithms' performance on a test set (untuned): matrix imputation (collaborative filtering)}
\label{FigMPad}
\end{figure*}

\begin{figure*}[b]
\centering
\includegraphics[width=16cm]{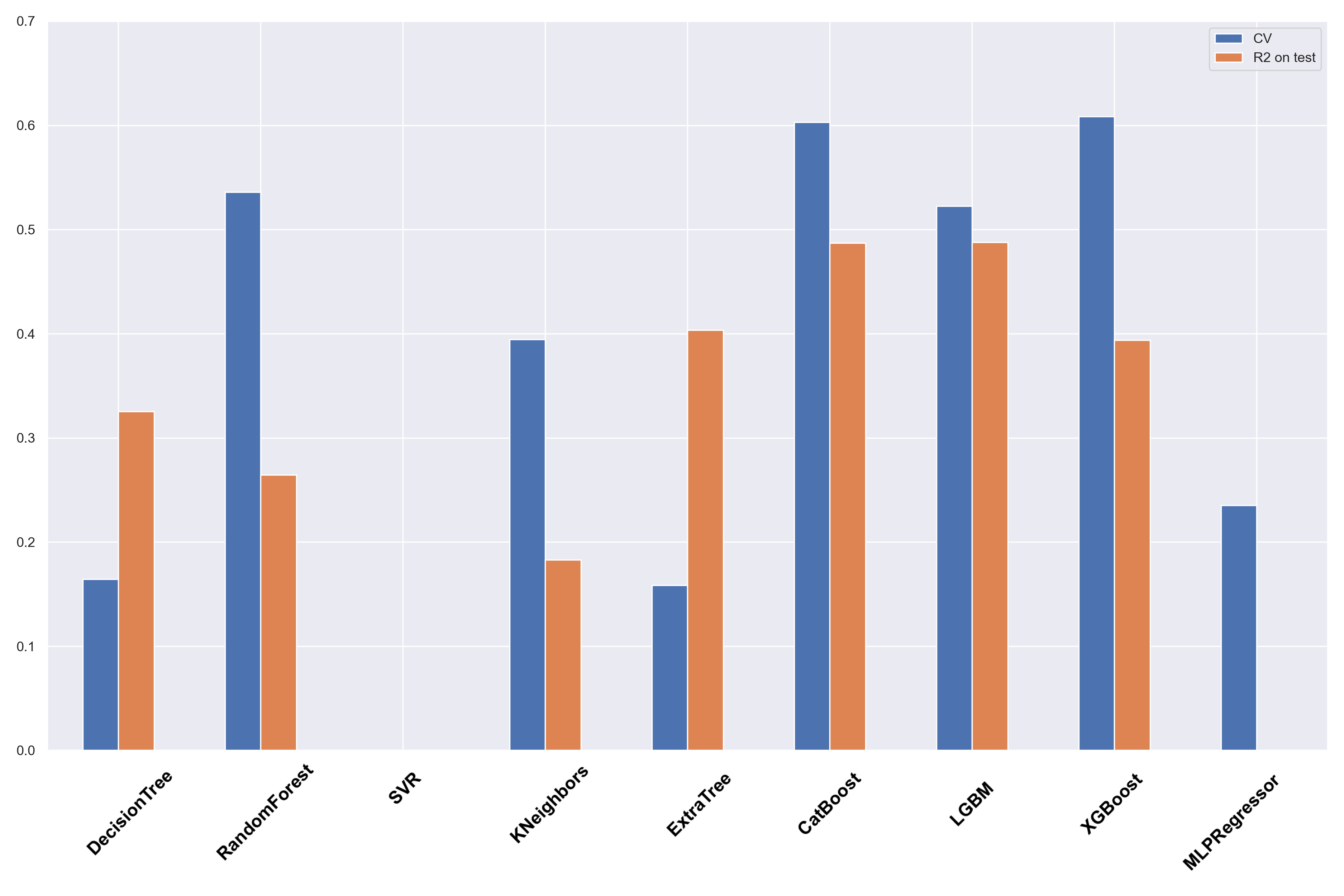}
\caption{Algorithms' performance on a test set (untuned): filling by mean values, calculated in each corresponding cluster}
\label{FigPClusters}
\end{figure*}

\end{document}